# Folded Codes from Function Field Towers and Improved Optimal Rate List Decoding[*]

Venkatesan Guruswami[†]     Chaoping Xing[‡]

April 2012

## Abstract

We give a new construction of algebraic codes which are efficiently list decodable from a fraction $1 - R - \varepsilon$ of adversarial errors where $R$ is the rate of the code, for any desired positive constant $\varepsilon$. The worst-case list size output by the algorithm is $O(1/\varepsilon)$, matching the existential bound for random codes up to constant factors. Further, the alphabet size of the codes is a constant depending only on $\varepsilon$ — it can be made $\exp(\tilde{O}(1/\varepsilon^2))$ which is not much worse than the lower bound of $\exp(\Omega(1/\varepsilon))$. The parameters we achieve are thus quite close to the existential bounds in all three aspects — error-correction radius, alphabet size, and list-size — simultaneously. Our code construction is Monte Carlo and has the claimed list decoding property with high probability. Once the code is (efficiently) sampled, the encoding/decoding algorithms are deterministic with a running time $O_\varepsilon(N^c)$ for an absolute constant $c$, where $N$ is the code's block length.

Our construction is based on a linear-algebraic approach to list decoding folded codes from towers of function fields, and combining it with a special form of subspace-evasive sets. Instantiating this with the explicit "asymptotically good" Garcia-Stichtenoth tower of function fields yields the above parameters. To illustrate the method in a simpler setting, we also present a construction based on Hermitian function fields, which offers similar guarantees with a list and alphabet size polylogarithmic in the block length $N$. Along the way, we shed light on how to use automorphisms of certain function fields to enable list decoding of the folded version of the associated algebraic-geometric codes.

---

[*]An extended abstract will appear in the *Proceedings of the 44th ACM Symposium on Theory of Computing* (STOC), 2012.

[†]Computer Science Department, Carnegie Mellon University, Pittsburgh, USA. `guruswami@cmu.edu`. Research supported in part by a Packard Fellowship and NSF CCF 0963975. Some of this work was done during a visit to Nanyang Technological University. Any opinions, findings, and conclusions or recommendations expressed in this material are those of the author(s) and do not necessarily reflect the views of the National Science Foundation.

[‡]Division of Mathematical Sciences, School of Physical & Mathematical Sciences, Nanyang Technological University, Singapore. `xingcp@ntu.edu.sg`. Research supported by the Singapore National Research Foundation under Research Grant NRF-CRP2-2007-03 and Singapore A*STAR SERC under Research Grant 1121720011.

# 1 Introduction

An error-correcting code $C$ of block length $N$ over a finite alphabet $\Sigma$ maps a set $\mathcal{M}$ of messages into codewords in $\Sigma^N$. The rate of the code $C$, denoted $R$, equals $\frac{1}{N}\log_{|\Sigma|}|\mathcal{M}|$. In this work, we will be interested in codes for adversarial noise, where the channel can arbitrarily corrupt any subset of up to $\tau N$ symbols of the codeword. The goal will be to correct such errors and recover the original message/codeword efficiently. It is easy to see that information-theoretically, we need to receive at least $RN$ symbols correctly in order to recover the message (since $|\mathcal{M}| = |\Sigma|^{RN}$), so we must have $\tau \leqslant 1 - R$.

Perhaps surprisingly, in a model called list decoding, recovery up to this information-theoretic limit becomes possible. Let us say that a code $C \subseteq \Sigma^N$ is $(\tau, \ell)$-list decodable if for every received word $\mathbf{y} \in \Sigma^N$, there are at most $\ell$ codewords $\mathbf{c} \in C$ such that $\mathbf{y}$ and $\mathbf{c}$ differ in at most $\tau N$ positions. Such a code allows, in principle, the correction of a fraction $\tau$ of errors, outputting at most $\ell$ candidate codewords one of which is the originally transmitted codeword.

The probabilistic method shows that a random code of rate $R$ over an alphabet of size $\exp(O(1/\varepsilon))$ is with high probability $(1-R-\varepsilon, O(1/\varepsilon))$-list decodable [Eli91]. However, it is not known how to construct or even randomly sample such a code for which the associated algorithmic task of list decoding (i.e., given $\mathbf{y} \in \Sigma^N$, find the list of codewords within fractional radius $1-R-\varepsilon$) can be performed efficiently. This work takes a big step in that direction, giving a randomized construction of such efficiently list-decodable codes over a slightly worse alphabet size of $\exp(\tilde{O}(1/\varepsilon^2))$. We note that the alphabet size needs to be at least $\exp(\Omega(1/\varepsilon))$ in order to list decode from a fraction $1-R-\varepsilon$ of errors, so this is close to optimal. For the list-size needed as a function of $\varepsilon$ for decoding a $1-R-\varepsilon$ fraction of errors, the best lower bound is only $\Omega(\log(1/\varepsilon))$ [GN12], but as mentioned above, even random coding arguments only achieve a list-size of $O(1/\varepsilon)$, which our construction matches up to constant factors.

We now review some of the key results on algebraic list decoding leading up to this work. A more technical comparison with related work appears in Section 1.1. The first construction of codes that achieved the optimal trade-off between rate and list-decoding radius, i.e., enabled list decoding up to a fraction $1-R-\varepsilon$ of worst-case errors with rate $R$, was due to Guruswami and Rudra [GR08]. They showed that a variant of Reed-Solomon (RS) codes called *folded* RS codes admit such a list decoder. For a decoding radius of $1-R-\varepsilon$, the code was based on bundling together disjoint windows of $m = \Theta(1/\varepsilon^2)$ consecutive symbols of the RS codeword into a single symbol over a larger alphabet. As a result, the alphabet size of the construction was $N^{\Omega(1/\varepsilon^2)}$. Ideas based on code concatenation and expander codes can be used to bring down the alphabet size to $\exp(\tilde{O}(1/\varepsilon^4))$, but this compromises some nice features such as list recovery and soft decoding of the folded RS code. Also, the decoding time complexity as well as proven bound on worst-case output list size for these constructions were $N^{\Omega(1/\varepsilon)}$ which is rather large.

Our main final result statement is the following. The codes we construct are a randomly sampled subcode of an analog of folded Reed-Solomon codes for an asymptotically optimal tower of function fields due to Garcia and Stichtenoth [GS95, GS96].

**Theorem 1.1** (Main). *For any $R \in (0,1)$ and positive constant $\varepsilon \in (0,1)$, there is a Monte Carlo construction of a family of codes of rate at least $R$ over an alphabet size $\exp(O(\log(1/\varepsilon)/\varepsilon^2))$ that are encodable and $(1-R-\varepsilon, O(1/(R\varepsilon)))$-list decodable in $O_\varepsilon(N^c)$ time, where $N$ is the block length* *of the code and c is an absolute positive constant.*

Even though our codes are not fully explicit, they are "functionally explicit" in the sense that once the code is (efficiently) sampled, with high probability the polynomial time encoding and decoding algorithms deliver the claimed error-correction guarantees for *all* allowed error patterns. We note that our codes are quite close to the existential bounds in three aspects simultaneously — the trade-off between error fraction $1 - R - \varepsilon$ and rate $R$, the list-size as a function of $\varepsilon$, and the alphabet size of the code family (again as a function of $\varepsilon$). Our algorithms can also be extended to the "list recovery" setting in a manner similar to [GR08, Gur11]; we omit discussion of this aspect and the straightforward details.

To first illustrate our ideas in an algebraically simpler (and perhaps more practical) setting, we first give a construction based on a tower of Hermitian field extensions [She93]. This gives a similar result, albeit with alphabet size and list-size upper bound polylogarithmic in $N$.

## 1.1 Prior and related work

Let us recap a bit more formally the construction of folded RS codes from [GR08]. One begins with the Reed-Solomon encoding of a polynomial $f \in \mathbb{F}_q[X]$ of degree $< k$ consisting of the evaluation of $f$ on a subset of field elements ordered as $1, \gamma, \ldots, \gamma^{n-1}$ for some primitive element $\gamma \in \mathbb{F}_q$ and $n < q$. For an integer "folding" parameter $m \geqslant 1$ that divides $n$, the folded RS codeword is defined over alphabet $\mathbb{F}_q^m$ and consists of $n/m$ blocks, with the $j$'th block consisting of the $m$-tuple $(f(\gamma^{(j-1)m}), f(\gamma^{(j-1)m+1}), \ldots, f(\gamma^{jm-1}))$. The algorithm in [GR08] for list decoding these codes was based on the algebraic identity $\overline{f(\gamma X)} = \overline{f(X)}^q$ in the residue field $\mathbb{F}_q[X]/(X^{q-1} - \gamma)$ where $\overline{f}$ denotes the residue $f \mod (X^{q-1} - \gamma)$. This identity is used to solve for $f$ from an equation of the form $Q(X, f(X), f(\gamma X), \ldots, f(\gamma^{s-1} X)) = 0$ for some low-degree nonzero multivariate polynomial $Q$. The high degree $q > n$ of this identity, coupled with $s \approx 1/\varepsilon$, led to the large bounds on list-size and decoding complexity in [GR08].

One possible approach to reduce $q$ (as a function of the code length) in this construction would be to work with algebraic-geometric codes based on function fields $K$ over $\mathbb{F}_q$ with more rational points. However, an automorphism $\sigma$ of $K$ that can play the role of the automorphism $f(X) \mapsto f(\gamma X)$ of $\mathbb{F}_q(X)$ is only known (or even possible) for very special function fields. This approach was used in [Gur10] to construct list-decodable codes based on cyclotomic function fields using as $\sigma$ certain Frobenius automorphisms. These codes improved the alphabet size to polylogarithmic in $N$, but the bound on list-size and decoding complexity remained $N^{\Omega(1/\varepsilon)}$.

Recently, a linear-algebraic approach to list decoding folded RS codes was discovered in [Vad10, Gur11]. Here, in the interpolation stage, which is common to all list decoding algorithms for algebraic codes [Sud97, GS99, PV05, GR08], following the idea in [Vad10] one finds a *linear* multivariate polynomial $Q(X, Y_1, \ldots, Y_s)$ whose total degree in the $Y_i$'s is 1. The simple but key observation driving [Gur11] is that the equation $Q(X, f(X), \ldots, f(\gamma^{s-1} X)) = 0$ now becomes a linear system in the coefficients of $f$. Further, it is shown that the solution space has dimension less than $s$, which again gives a list-size upper bound of $q^{s-1}$. Finally, since the list of candidate messages fall in an affine space, it was noted in [Gur11] that one can bring down the list size by carefully "pre-coding" the message polynomials so that their $k$ coefficients belong to a "subspace-evasive set" (which has small intersection with every $s$-dimensional subspace of $\mathbb{F}_q^k$). This idea was used in [Gur11] to give a *randomized* construction of $(1 - R - \varepsilon, O(1/\varepsilon^2))$-list decodable codes of rate $R$ (in fact, the list size bound is worse — it is $\tilde{\Omega}(N)$ — if one requires efficient encoding of the code). However, the alphabet size and runtime of the decoding algorithm both remained $N^{\Omega(1/\varepsilon)}$. In [GW11], similar results were also shown for derivative codes, where the encoding of a polynomial $f$ consists of the evaluations of $f$ and its first $m-1$ derivates at distinct field elements.

In a concurrent independent work, Dvir and Lovett gave an elegant construction of explicit subspace evasive sets based on certain algebraic varieties [DL11]. This yields an explicit version of the codes from [Gur11], albeit with a worse list size bound of $(1/\varepsilon)^{O(1/\varepsilon)}$. This work and [DL11] are incomparable in terms of results. The big advantage of [DL11] is the deterministic construction of the code. The benefits in our work are (i) our list-size of $O(1/\varepsilon)$ is much better and in fact optimal up to constant factors,[1] and (ii) we are able to construct codes over an alphabet size that is a constant independent of $N$, whereas in [DL11] the $N^{\Omega(1/\varepsilon^2)}$ alphabet size of folded RS codes is inherited. Both our work and [DL11] achieve a decoding complexity of $O_\varepsilon(N^c)$ with exponent independent of $\varepsilon$.

We should note that since we require sets that are evasive with respect to subspaces of large dimension, and which have further structural properties needed in the decoding, we cannot use the construction in [DL11] to make the codes in this work explicit.

## 1.2 Our techniques

We describe some of the main new ingredients that go into our work. We need both new algebraic insights and constructions, as well as ideas in pseudorandomness relating to subspace-evasive sets with additional structure. We describe these in turn below.

**Algebraic ideas.** As mentioned above, effecting the original "non-linear" approach in [GR08, Gur10] with automorphisms of more general function fields seems intricate at best. The correct generalization of the linear-algebraic list decoding approach to the function field case is also not obvious. One of the main algebraic insights in this work is noting that the right way to generalize the linear-algebraic approach to codes based on algebraic function fields is to rely on the *local power series expansion* of functions from the message space at a suitable rational point. (The case for Reed-Solomon codes being the expansion around 0, which is a finite polynomial form.)

Working with a suitable automorphism which has a "diagonal" action on the local expansion lets us extend the linear-algebraic decoding method to AG codes. Implementing this for specific AG codes requires an explicit specification of a basis for an associated message (Riemann-Roch) space, and the efficient computation of the local expansion of the basis elements at a special rational point on the curve. We show how to do this for two towers of function fields: the Hermitian tower [She93] and the asymptotically optimal Garcia-Stichtenoth tower [GS95, GS96]. The former tower is quite simple to handle — it has an easily written down explicit basis, and we show how to compute the local expansion of functions around the point with all zero coordinates. However, the Hermitian tower does not have bounded ratio of the genus to number of rational points, and so does not give constant alphabet codes (we can get codes over an alphabet size that is polylogarithmic in the block length though). Explicit basis for Riemann-Roch spaces of the Garcia-Stichtenoth tower were constructed in [SAK+01]. Regarding local expansions, one major difference is that we work with local expansion of functions at the point at infinity, which is fully "ramified" in the tower. For both these towers, we find and work with a nice automorphism that acts diagonally on the local expansion, and use it for folding the codes and decoding them by solving a linear system.

---
[1]As mentioned above, the bound in [DL11] is $(1/\varepsilon)^{O(1/\varepsilon)}$ and it seems very difficult to get a sub-exponential dependence on $1/\varepsilon$ with the algebraic approach relying on Bezout's theorem to construct subspace-evasive sets.

**Pseudorandomness.** These algebraic ideas enable us to pin down the messages into a subspace of dimension linear in the message length. To prune this list, we need several additional ideas. The starting point is to follow [Gur11] and only encode messages in a subspace-evasive set which has small intersection with low-dimensional subspaces. Implementing this in our case, however, leads to several problems. First, since the subspace we like to avoid intersecting much has large dimension, the list size bound will be linear in the code length and not a constant like in our final result. More severely, we cannot go over the elements of this subspace to prune the list as that would take exponential time. To solve the latter problem, we observe that the subspace has a special "periodic" structure, and exploit this to show the existence of large "hierarchically subspace evasive" (h.s.e) subsets which have small intersection with the projection of the subspace on certain prefixes. Isolating the periodic property of the subspaces, and formulating the right notion of evasiveness w.r.t to such subspaces, is an important aspect of this work.

We also give a pseudorandom construction of good h.s.e sets using limited wise independent sample spaces, in a manner enabling the efficient iterative computation of the final list of intersecting elements. With some additional ideas, we ensure that one can efficiently index into a large subset of our h.s.e set construction (this is needed to get an efficient encoding algorithm for our code). As a further ingredient, we note that the number of possible subspaces that arise in the decoding is much smaller than the total number of possibilities. Using this together with a trick to take the intersection of two subspace evasive set constructions, we are able to reduce the list size to a constant.

## 1.3 Organization

We begin by isolating the special notion of subspaces which our evasive sets should avoid intersecting too much (Section 2). We describe our construction of folded Hermitian codes and a linear-algebraic list decoding algorithm for these codes in Section 3. In Section 4, we define and construct the special "hierarchically" subspace-evasive (h.s.e) sets that we need. We show how to combine the h.s.e sets with folded Hermitian codes in Section 5; this gives a result similar to Theorem 1.1 with polylogarithmic alphabet and list size. We show how our ideas can be used to construct folded codes based on the Garcia-Stichtenoth tower, and how to combine them with h.s.e sets to get our main result (Theorem 1.1) in Section 6.

## 2 Periodic subspaces

The list decoding algorithm for our algebraic codes will first pin down the candidate messages to a subspace. The structure of the subspace will be important to us in order to be able to efficiently prune it to a much smaller list. In this section, we make some important definitions capturing this property. Let us begin with some notation.

**Notation** (Projection of vectors and sets). *For a vector* $\mathbf{y} = (y_1, y_2, \ldots, y_m) \in \mathbb{F}_q^m$ *and positive*

integers $t_1 \leqslant t_2 \leqslant m$, we denote by $\mathrm{proj}_{[t_1,t_2]}(\mathbf{y}) \in \mathbb{F}_q^{t_2-t_1+1}$ its projection onto coordinates $t_1$ through $t_2$, i.e., $\mathrm{proj}_{[t_1,t_2]}(\mathbf{y}) = (y_{t_1}, y_{t_1+1}, \ldots, y_{t_2})$. When $t_1 = 1$, we use $\mathrm{proj}_t(\mathbf{y})$ to denote $\mathrm{proj}_{[1,t]}(\mathbf{y})$.

For a subset $S \subseteq \mathbb{F}_q^k$ and positive integers $t_1 \leqslant t_2 \leqslant k$, we denote the projection of $S$ onto the coordinates in the range $[t_1, t_2]$ by $\mathrm{proj}_{[t_1,t_2]}(S)$. Formally, $\mathrm{proj}_{[t_1,t_2]}(S) = \{\mathrm{proj}_{[t_1,t_2]}(\mathbf{y}) \mid \mathbf{y} \in S\}$. Again, we use $\mathrm{proj}_t(S)$ to denote $\mathrm{proj}_{[1,t]}(S)$.

The specific definition of periodic subspaces below (which might appear rather technical) is motivated by the structure of the subspaces arising in our list decoding application (for example, as guaranteed by Lemma 3.7). The special structure of these subspaces is important to guarantee the existence of "subspace evasive sets" (defined later) that are good enough for our purposes.

**Definition 1.** $(s, \Delta)$**-periodic subspaces**[2]) *For positive integers $s, \Delta, b$, an affine subspace $W$ of $\mathbb{F}_q^k$ where $k = b\Delta$ is said to be $(s, \Delta)$-periodic if there is a subspace $U$ of $\mathbb{F}_q^\Delta$ of dimension less than $s$ such that for all $\mathbf{y} \in W$ and $1 \leqslant i \leqslant b$, $\mathrm{proj}_{[(i-1)\Delta+1, i\Delta]}(\mathbf{y})$ belongs to the affine space $U + \mathbf{b}_i$, where $\mathbf{b}_i$ is a column vector whose coordinates are affine combinations (depending only on $i$) of the first $(i-1)\Delta$ coordinates of $\mathbf{y}$; formally, $\mathbf{b}_i = C_i \cdot \mathrm{proj}_{(i-1)\Delta}(\mathbf{y}) + \mathbf{v}_i$ for some matrix $C_i \in \mathbb{F}_q^{\Delta \times (i-1)\Delta}$ and $\mathbf{v}_i \in \mathbb{F}_q^\Delta$. We can represent such an affine subspace by $U, \{C_i, \mathbf{v}_i\}_{i=1}^b$.*

Note that if $W$ is an $(s, \Delta)$-periodic subspace of $\mathbb{F}_q^{b\Delta}$, for every $i$, $1 \leqslant i \leqslant b$, and every $\mathbf{a} \in \mathbb{F}_q^{(i-1)\Delta}$, the affine space $\{\mathrm{proj}_{[(i-1)\Delta+1, i\Delta]}(\mathbf{w}) \mid \mathbf{w} \in W \text{ and } \mathrm{proj}_{(i-1)\Delta}(\mathbf{w}) = \mathbf{a}\}$ has dimension at most $s$ (and in particular it has at most $q^s$ elements). Therefore, by an inductive argument, we have $|\mathrm{proj}_{i\Delta}(W)| \leqslant q^{is}$ for $1 \leqslant i \leqslant b$, which together with the fact that $\mathrm{proj}_{i\Delta}(W)$ is an affine subspace implies the following.

**Observation 2.1.** *If $W$ is an $(s, \Delta)$-periodic subspace of $\mathbb{F}_q^{b\Delta}$, then for $i = 1, 2, \ldots, b$, $\mathrm{proj}_{i\Delta}(W)$ is also $(s, \Delta)$-periodic and has dimension at most $s \cdot i$ as an affine subspace of $\mathbb{F}_q^{i\Delta}$.*

## 3 Folded codes from the Hermitian tower

In this section, we will describe a family of folded codes based on the Hermitian function field (or rather a tower of such fields).

### 3.1 Background on Hermitian tower

In what follows, let $r$ be a prime power and let $q = r^2$. We denote by $\mathbb{F}_q$ the finite field with $q$ elements. The Hermitian function tower that we are going to use for our code construction was discussed in [She93]. The reader may refer to [She93] for the detailed background on the Hermitian function tower, and Stichtenoth's book [Sti93] for general background on algebraic function fields and their use in constructing algebraic-geometric codes. The Hermitian tower is defined by the following recursive equations

$$x_{i+1}^r + x_{i+1} = x_i^{r+1}, \quad i = 1, 2, \ldots, e-1.$$

---

[2] According to the definition, an $(s, \Delta)$-periodic subspace is in fact an affine space. For convenience, we blur this distinction, which is not too important for us, and use the terminology periodic subspace to refer to them.

Put $F_e = \mathbb{F}_q(x_1, x_2, \ldots, x_e)$ for $e \geq 2$. We will assume that $r \geq 2e$.

**Rational places.** The function field $F_e$ has $r^{e+1} + 1$ rational places. One of these is the "point at infinity" which is the unique pole $P_\infty$ of $x_1$ (and is fully ramified). The other $r^{e+1}$ come from the rational places lying over the unique zero $P_\alpha$ of $x_1 - \alpha$ for each $\alpha \in \mathbb{F}_q$. Note that for every $\alpha \in \mathbb{F}_q$, $P_\alpha$ splits completely in $F_e$, i.e., there are $r^{e-1}$ rational places lying over $P_\alpha$. Intuitively, one can think of the rational places of $F_e$ (besides $P_\infty$) as being given by $e$-tuples $(\alpha_1, \alpha_2, \ldots, \alpha_e) \in \mathbb{F}_q^e$ that satisfy $\alpha_{i+1}^r + \alpha_{i+1} = \alpha_i^{r+1}$ for $i = 1, 2, \ldots, e-1$. For each value of $\alpha \in \mathbb{F}_q$, there are precisely $r$ solutions to $\beta \in \mathbb{F}_q$ satisfying $\beta^r + \beta = \alpha^{r+1}$, so the number of such $e$-tuples is $r^{e+1}$ ($q = r^2$ choices for $\alpha_1$, and then $r$ choices for each successive $\alpha_i$, $2 \leq i \leq e$).

**Riemann-Roch spaces.** For a place $P$ of $F_e$, we denote by $\nu_P$ the discrete valuation of $P$: for a function $h \in F_e$, if $h$ has a zero at $P$, then $\nu_P(h)$ gives the number (multiplicity) of zeroes, if $h$ has a pole at $P$, then $-\nu_P(h)$ gives the pole order of $h$ at $P$, and $\nu_P(h) = 0$ if $h$ has neither a zero or a pole at $P$.

For an integer $l$, we consider the Riemann-Roch space defined by
$$\mathcal{L}(lP_\infty) := \{h \in F_e \setminus \{0\} : \nu_{P_\infty}(h) \geq -l\} \cup \{0\}.$$

Then the dimension $\ell(lP_\infty)$ is at least $l - g_e + 1$ and furthermore, $\ell(lP_\infty) = l - g_e + 1$ if $l \geq 2g_e - 1$. A basis over $\mathbb{F}_q$ of $\mathcal{L}(lP_\infty)$ can be explicitly constructed as follows
$$\left\{ x_1^{j_1} \cdots x_e^{j_e} : (j_1, \ldots, j_e) \in \mathbb{Z}_{\geq 0}^e, \ \sum_{i=1}^e j_i r^{e-i}(r+1)^{i-1} \leq l \right\}. \tag{1}$$

We stress that evaluating elements of $\mathcal{L}(lP_\infty)$ at the rational places of $F_e$ (other then $P_\infty$) is easy: we simply have to evaluate a linear combination of the monomials allowed in (1) at the tuples $(\alpha_1, \alpha_2, \ldots, \alpha_e) \in \mathbb{F}_q^e$ mentioned above. In other words, it is just evaluating an $e$-variate polynomial at a specific subset of $r^{e+1}$ points of $\mathbb{F}_q^e$, and can be accomplished in polynomial time.

**Genus.** The genus $g_e$ of the function field $F_e$ is given by
$$g_e = \frac{1}{2}\left( \sum_{i=1}^{e-1} r^e \left(1 + \frac{1}{r}\right)^{i-1} - (r+1)^{e-1} + 1 \right) \leq \frac{r^e}{2} \sum_{i=1}^{e} \binom{e}{i} \frac{1}{r^{i-1}} \leq \frac{er^e}{2} \sum_{i=1}^{e} \left(\frac{e}{r}\right)^{i-1} \leq er^e \tag{2}$$

where the last step used $r \geq 2e$.

**A useful automorphism.** Let $\gamma$ be a primitive element of $\mathbb{F}_q$ and consider the automorphism $\sigma \in \text{Aut}(F_e/\mathbb{F}_q)$ defined by
$$\sigma : x_i \mapsto \gamma^{(r+1)^{i-1}} x_i \quad \text{for } i = 1, 2, \ldots, e.$$

The order of $\sigma$ is $q - 1$ and furthermore, we have the following facts:

(i) Let $P_0$ be the unique common zero of $x_1, x_2, \ldots, x_e$ (this corresponds to the $e$-tuple $(0, 0, \ldots, 0)$), and $P_\infty$ the unique pole of $x_1$. The automorphism $\sigma$ keeps $P_0$ and $P_\infty$ unchanged, i.e., $P_0^\sigma = P_0$ and $P_\infty^\sigma = P_\infty$,

(ii) Let $\mathbb{P}$ be the set of all the rational places which are neither $P_\infty$ nor zeros of $x_1$. Then $|\mathbb{P}| = (q-1)r^{e-1}$. Moreover, $\sigma$ divides $\mathbb{P}$ into $r^{e-1}$ orbits and each orbit has $q-1$ places. For an integer $m$ with $1 \leqslant m \leqslant q-1$, we can label $Nm$ distinct elements $P_1, P_1^\sigma, \ldots, P_1^{\sigma^{m-1}}, \ldots, P_N, P_N^\sigma, \ldots, P_N^{\sigma^{m-1}}$ in $\mathbb{P}$, as long as $N \leqslant r^{e-1} \left\lfloor \frac{q-1}{m} \right\rfloor$.

**Definition 2** (Folded codes from the Hermitian tower). *Assume that $m, l, N$ are positive integers satisfying $1 \leqslant m \leqslant q-1$ and $l/m \leqslant N \leqslant r^{e-1} \left\lfloor \frac{q-1}{m} \right\rfloor$. The folded code from $F_e$ with parameters $N, l, q, e, m$, denoted by $\widetilde{\mathsf{FH}}(N, l, q, e, m)$, encodes a message function $f \in \mathcal{L}(lP_\infty)$ as*

$$f \mapsto \left( \begin{bmatrix} f(P_1) \\ f(P_1^\sigma) \\ \vdots \\ f(P_1^{\sigma^{m-1}}) \end{bmatrix}, \begin{bmatrix} f(P_2) \\ f(P_2^\sigma) \\ \vdots \\ f(P_2^{\sigma^{m-1}}) \end{bmatrix}, \ldots, \begin{bmatrix} f(P_N) \\ f(P_N^\sigma) \\ \vdots \\ f(P_N^{\sigma^{m-1}}) \end{bmatrix} \right) \in \left( \mathbb{F}_q^m \right)^N. \tag{3}$$

**Lemma 3.1.** *The above code $\widetilde{\mathsf{FH}}(N, l, q, e, m)$ is an $\mathbb{F}_q$-linear code over alphabet size $q^m$, rate at least $\frac{l - g_e + 1}{Nm}$, and minimum distance at least $N - \frac{l}{m}$.*

*Proof.* It is clear that the map (3) is an $\mathbb{F}_q$-linear map. The dimension over $\mathbb{F}_q$ of the message space $\mathcal{L}(lP_\infty)$ is at least $l - g_e + 1$ by the Riemann-Roch theorem, which gives the claimed lower bound on rate. For the distance property, observe that if the $i$-th column is zero, then $f$ has $m$ zeros. This implies that the encoding of a nonzero function $f$ can have at most $l/m$ zero columns since $f \in \mathcal{L}(lP_\infty)$. □

## 3.2 Redefining the code in terms of local expansion at $P_0$

For our decoding, we will actually recover the message $f \in \mathcal{L}(lP_\infty)$ in terms of the coefficients of its power series expansion around $P_0$

$$f = f_0 + f_1 x + f_2 x^2 + \cdots$$

where $x := x_1$ is the local parameter at $P_0$ (which means that $x_1$ has exactly one zero at $P_0$, i.e., $\nu_{P_0}(x_1) = 1$). In fact, realizing that one must work in this power series representation is one of the key insights in this work.

Let us first show that one can efficiently move back-and-forth between the representation of $f \in \mathcal{L}(lP_\infty)$ in terms of a basis for $\mathcal{L}(lP_\infty)$ and its power series representation $(f_0, f_1, \ldots)$ around $P_0$. Since the mapping $f \mapsto (f_0, f_1, \ldots)$ is $\mathbb{F}_q$-linear, it suffices to compute the local expansion at $P_0$ of a basis for $\mathcal{L}(lP_\infty)$.

**Lemma 3.2.** *For any $n$, one can compute the first $n$ terms of the local expansion of the basis elements (1) at $P_0$ using $\mathrm{poly}(n)$ operations over $\mathbb{F}_q$.*

*Proof.* By the structure of the basis functions in (1), it is sufficient to find an algorithm of efficiently finding local expansions of $x_i$ at $P_0$ for every $i = 1, 2, \ldots, e$. We can inductively find the local expansions of $x_i$ at $P_0$ as follows.

For $i = 1$, $x_1$ is the local parameter $x$ of $P_0$, so $x$ is the local expansion of $x_1$ at $P_0$.

Now assume that we know the local expansion of $x_i = \sum_{j=1}^{\infty} c_{i,j} x^j$ at $P_0$ for some $c_{i,j} \in \mathbb{F}_q$. Then we have

$$\sum_{j=1}^{\infty} c_{i+1,j}^r x^{jr} + \sum_{j=1}^{\infty} c_{i+1,j} x^j = x_{i+1}^r + x_{i+1} = x_i^{r+1} = \left(\sum_{j=1}^{\infty} c_{i,j}^r x^{jr}\right)\left(\sum_{j=1}^{\infty} c_{i,j} x^j\right).$$

By comparing the coefficients of $x^j$ in the above identity, we can easily solve $c_{i+1,j}$'s from $c_{i,j}$'s. More specifically, the coefficient of $x^j$ at the left of the identity is

$$\begin{cases} c_{i+1,j} & \text{if } r \nmid j \\ c_{i+1,j} + c_{i+1,j/r}^r & \text{if } r \mid j. \end{cases}$$

Thus, all $c_{i+1,j}$'s can be easily solved recursively. $\square$

To keep the list output by the algorithm at a controllable size, we will combine the code with certain special subspace evasive sets. For this purpose, we will actually need to index the messages of the code by the first $k$ coefficients $(f_0, f_1, \ldots, f_{k-1})$ of the local expansion of the function $f$ at $P_0$. This requires that for every $(f_0, f_1, \ldots, f_{k-1})$ there is a $f \in \mathcal{L}(lP_\infty)$ whose power series expansion has the $f_i$ as the first $k$ coefficients. This is easy to ensure by taking $l = k + 2g_e - 1$ as we argue below. Note that to ensure that $\mathcal{L}(lP_\infty)$ has dimension $k$, it suffices to pick $l = k + g_e - 1$ by the Riemann-Roch theorem. We pick $l$ to be $g_e$ more than this bound. Since the genus will be much smaller than the code length, we can afford this small loss in parameters.

Let us define the local expansion map $\mathsf{ev}_{P_0} : \mathcal{L}((k + 2g_e - 1)P_\infty) \to \mathbb{F}_q^k$ that maps $f$ to $(f_0, f_1, \ldots, f_{k-1})$ where $f = f_0 + f_1 x + f_2 x^2 + \cdots$ is the local expansion of $f$ at $P_0$.

**Claim 3.3.** $\mathsf{ev}_{P_0}$ *is an $\mathbb{F}_q$-linear surjective map. Further, we can compute $\mathsf{ev}_{P_0}$ using $\mathrm{poly}(k, g_e)$ operations over $\mathbb{F}_q$ given a representation of the input $f \in \mathcal{L}((k+2g_e-1)P_\infty)$ in terms of the basis* (1).

*Proof.* The $\mathbb{F}_q$-linearity of $\mathsf{ev}_{P_0}$ is clear. The kernel of $\mathsf{ev}_{P_0}$ is $\mathcal{L}((k + 2g_e - 1)P_\infty - kP_0)$ which has dimension exactly $g_e$ by the Riemann-Roch theorem. By the rank-nullity theorem, the image must have dimension $k$, and so the map is surjective. The claimed complexity of computation follows immediately from Lemma 3.2. $\square$

For each $(f_0, f_1, \ldots, f_{k-1}) \in \mathbb{F}_q^k$, we can therefore pick a pre-image in $\mathcal{L}((k + 2g_e - 1)P_\infty)$. For convenience, we will denote an injective map making such a unique choice by $\kappa_{P_0} : \mathbb{F}_q^k \to \mathcal{L}((k + 2g_e - 1)P_\infty)$. By picking the pre-images of a basis of $\mathbb{F}_q^k$ and extending it by linearity, we can assume $\kappa_{P_0}$ to be $\mathbb{F}_q$-linear, and thus specify it by a $(k + g_e) \times k$ matrix. We record this fact for easy reference below.

**Claim 3.4.** *The map $\kappa_{P_0} : \mathbb{F}_q^k \to \mathcal{L}((k+2g_e-1)P_\infty)$ is $\mathbb{F}_q$-linear and injective. We can compute a representation of this linear transformation using $\mathrm{poly}(k, g_e)$ operations over $\mathbb{F}_q$, and the map itself can be evaluated using $\mathrm{poly}(k, g_e)$ operations over $\mathbb{F}_q$.*

We will now redefine a version of the folded Hermitian code that maps $\mathbb{F}_q^k$ to $(\mathbb{F}_q^m)^N$ by composing the folded encoding (3) from the original Definition 2 with $\kappa_{P_0}$.

**Definition 3** (Folded Hermitian code using local expansion)**.** *The folded Hermitian code* $\mathsf{FH}(N, k, q, e, m)$ *maps* $\mathbf{f} = (f_0, f_1, \ldots, f_{k-1}) \in \mathbb{F}_q^k$ *to* $\widetilde{\mathsf{FH}}(N, k + 2g_e - 1, q, e, m)(\kappa_{P_0}(\mathbf{f})) \in (\mathbb{F}_q^m)^N$.

The rate of the above code equals $k/(Nm)$ and its distance is at least $N - (k + 2g_e - 1)/m$.

## 3.3 List decoding folded codes from the Hermitian tower

We now present a list decoding algorithm for the above codes. The algorithm follows the linear-algebraic list decoding algorithm for folded Reed-Solomon codes. Suppose a codeword (3) encoding $f \in \mathrm{Im}(\kappa_{P_0}) \subseteq \mathcal{L}((k + 2g_e - 1)P_\infty)$ is transmitted and received as

$$\mathbf{y} = \begin{pmatrix} y_{1,1} & y_{2,1} & & y_{N,1} \\ y_{1,2} & y_{2,2} & & \vdots \\ & & \ddots & \\ y_{1,m} & \cdots & & y_{N,m} \end{pmatrix} \tag{4}$$

where some columns are erroneous. Let $s \geq 1$ be an integer parameter associated with the decoder.

**Lemma 3.5.** *Given a received word as in (4), using* $\mathrm{poly}(N)$ *operations over* $\mathbb{F}_q$, *we can find a nonzero linear polynomial in* $F_e[Y_1, Y_2, \ldots, Y_s]$ *of the form*

$$Q(Y_1, Y_2, \ldots, Y_s) = A_0 + A_1 Y_1 + A_2 Y_2 + \cdots + A_s Y_s \tag{5}$$

*satisfying*

$$Q(y_{i,j}, y_{i,j+1}, \cdots, y_{i,j+s-1}) = A_0(P_i^{\sigma^j}) + A_s(P_i^{\sigma^j}) y_{i,j+1} + \cdots + A_s(P_i^{\sigma^j}) y_{i,j+s} = 0 \tag{6}$$

*for* $i = 1, 2, \ldots, N$ *and* $j = 0, 1, \ldots, m - s$. *The coefficients* $A_i$ *of* $Q$ *satisfy* $A_i \in \mathcal{L}(DP_\infty)$ *for* $i = 1, 2, \ldots, s$ *and* $A_0 \in \mathcal{L}((D + k + 2g_e - 1)P_\infty)$ *for a "degree" parameter* $D$ *chosen as*

$$D = \left\lfloor \frac{N(m - s + 1) - k + (s - 1)g_e + 1}{s + 1} \right\rfloor. \tag{7}$$

*Proof.* If we fix a basis of $\mathcal{L}(DP_\infty)$ (of the form (1)) and extend it to a basis of $\mathcal{L}((D+k+2g_e-1)P_\infty)$, then the number of freedoms of $A_0$ is at least $D + k + g_e$ and the number of freedoms of $A_i$ is at least $D - g_e + 1$ for $i \geq 1$. Thus, the total number of freedoms in the polynomial $Q$ equals

$$s(D - g_e + 1) + D + k + g_e = (s + 1)(D + 1) - (s - 1)g_e - 1 + k > N(m - s + 1) \tag{8}$$

for the above choice (7) of $D$. The interpolation requirements on $Q \in F_e[Y_1, \ldots, Y_s]$ are the following:

$$Q(y_{i,j}, y_{i,j+1}, \cdots, y_{i,j+s-1}) = A_0(P_i^{\sigma^j}) + A_s(P_i^{\sigma^j}) y_{i,j+1} + \cdots + A_s(P_i^{\sigma^j}) y_{i,j+s} = 0 \tag{9}$$

for $i = 1, 2, \ldots, N$ and $j = 0, 1, \ldots, m - s$. The interpolation requirements on $Q$ give a total of $N(m - s + 1)$ homogeneous linear equations that the coefficients of the $A_i$'s w.r.t the chosen basis of $\mathcal{L}((D + k + 2g_e - 1)P_\infty)$ must satisfy. Since the number of such coefficients (degrees of freedom in $Q$) exceeds $N(m - s + 1)$, we can conclude that such a linear polynomial $Q$ as required by the lemma must exist, and can be found by solving a homogeneous linear system over $\mathbb{F}_q$ with about $N(m - s + 1)$ variables and constraints. □

Similar to earlier interpolation based list decoding algorithms, the following lemma gives an algebraic condition that the message functions $f \in \mathcal{L}((k+2g_e-1)P_\infty)$ we are interested in list decoding must satisfy. The proof is a standard argument comparing the pole order to the number of zeroes.

**Lemma 3.6.** *If $f$ is a function in $\mathcal{L}((k+2g_e-1)P_\infty)$ whose encoding (3) agrees with the received word $\mathbf{y}$ in at least $t$ columns with $t > \frac{D+k+2g_e-1}{m-s+1}$, then*

$$Q(f, f^{\sigma^{-1}}, \ldots, f^{\sigma^{-(s-1)}}) = A_0 + A_1 f + A_2 f^{\sigma^{-1}} + \cdots + A_s f^{\sigma^{-(s-1)}} = 0. \tag{10}$$

*Proof.* The proof proceeds by comparing the number of zeros of the function $Q(f, f^{\sigma^{-1}}, \ldots, f^{\sigma^{-(s-1)}}) = A_0 + A_1 f + A_2 f^{\sigma^{-1}} + \cdots + A_s f^{\sigma^{-(s-1)}}$ with $D + k + 2g_e - 1$. Note that $Q(f, f^{\sigma^{-1}}, \ldots, f^{\sigma^{-(s-1)}})$ is a function in $\mathcal{L}((D+k+2g_e-1)P_\infty)$. If column $i$ of the encoding (3) of $f$ agrees with $\mathbf{y}$, then for all $j = 0, 1, \ldots, m-s$, we have

$$\begin{aligned}
0 &= A_0(P_i^{\sigma^j}) + A_1(P_i^{\sigma^j}) y_{i,j+1} + A_2(P_i^{\sigma^j}) y_{i,j+2} + \cdots + A_s(P_i^{\sigma^j}) y_{i,j+s} \\
&= A_0(P_i^{\sigma^j}) + A_1(P_i^{\sigma^j}) f(P_i^{\sigma^j}) + A_2(P_i^{\sigma^j}) f(P_i^{\sigma^{j+1}}) + \cdots + A_s(P_i^{\sigma^j}) f(P_i^{\sigma^{j+s-1}}) \\
&= A_0(P_i^{\sigma^j}) + A_1(P_i^{\sigma^j}) f(P_i^{\sigma^j}) + A_2(P_i^{\sigma^j}) f^{\sigma^{-1}}(P_i^{\sigma^j}) + \cdots + A_s(P_i^{\sigma^j}) f^{\sigma^{-(s-1)}}(P_i^{\sigma^j}) \\
&= (A_0 + A_1 f + A_2 f^{\sigma^{-1}} + \cdots + A_s f^{\sigma^{-(s-1)}})(P_i^{\sigma^j}) .
\end{aligned}$$

Note that here we use the fact that $f^\sigma(P^\sigma) = f(P)^\sigma = f(P)$, or equivalently $f(P^\sigma) = f^{\sigma^{-1}}(P)$. In other words, $Q(f, f^{\sigma^{-1}}, \ldots, f^{\sigma^{-(s-1)}})$ has $(m-s+1)$ distinct zeros from this agreeing column. Thus, there are a total of at least $t(m-s+1)$ zeros for all the agreeing columns. Hence, $Q(f, f^{\sigma^{-1}}, \ldots, f^{\sigma^{-(s-1)}})$ must be the zero function when $t(m-s-1) > D+k+2g_e-1$. □

***Solving the functional equation for $f$.*** Our goal next is to recover the list of solutions $f$ to the functional equation (10). Recall that our message functions lie in $\mathrm{Im}(\kappa_{P_0})$, so we can recover $f$ by recovering the top $k$ coefficients $(f_0, f_1, \ldots, f_{k-1})$ of its local expansion $f = \sum_{j=0}^\infty f_j x^j$ at $P_0$. We now prove that $(f_0, f_1, \ldots, f_{k-1})$ for $f$ satisfying Equation (10) belong to a "periodic" subspace (in the sense of Definition 1) of not too large dimension.

**Lemma 3.7.** *The set of solutions $(f_0, f_1, \ldots, f_{k-1}) \in \mathbb{F}_q^k$ such that $f = f_0 + f_1 x + f_2 x^2 + \cdots \in \mathcal{L}((k+2g_e-1)P_\infty)$ obeys equation*

$$A_0 + A_1 f + A_2 f^{\sigma^{-1}} + \cdots + A_s f^{\sigma^{-(s-1)}} = 0 , \tag{11}$$

*when the $A_i$'s obey the pole order restrictions of Lemma 3.5 and at least one $A_i$ is nonzero, is an affine subspace $W$ of dimension at most $(s-1) \left\lceil \frac{k}{q-1} \right\rceil$.*

*Further, there are at most $q^{Nm+s+1}$ possible choices of the subspace $W$ (as a function of the $A_i$'s), each of which is $(s, q-1)$-periodic. Given the representation of each $A_i$ w.r.t the basis (1), we can find a representation of $W$ in terms of the periodic subspace $U$ of dimension less than $s$, and the affine shifts in each window of $q-1$ coordinates, in the sense of Definition 1.*

*Proof.* Let $u = \min\{\nu_{P_0}(A_i) : i = 1, 2, \ldots, s\}$. Then it is clear that $u \geqslant 0$ and $\nu_{P_0}(A_0) \geqslant u$. Each $A_i$ has a local expansion at $P_0$:

$$A_i = x^u \sum_{j=0}^\infty a_{i,j} x^j$$

for $i = 0, 1, \ldots, s - 1$, which can be efficiently computed from the basis representation of the $A_i$'s. From the definition of $u$, one knows that the polynomial

$$B_0(X) := a_{1,0} + a_{2,0}X + \cdots + a_{s,0}X^{s-1}$$

is nonzero. Assume that at $P_0$, the function $f$ has a local expansion $\sum_{j=0}^{\infty} f_j x^j$. Then $f^{\sigma^{-i}}$ has a local expansion at $P_0$ as follows

$$f^{\sigma^{-i}} = \sum_{j=0}^{\infty} \xi^{ij} f_j x^j,$$

where $\xi = 1/\gamma$. The coefficient of $x^{d+u}$ in the local expansion of $Q(f, f^{\sigma^{-1}}, \ldots, f^{\sigma^{-(s-1)}})$ is

$$0 = B_0(\xi^d) f_d + \sum_{i=0}^{d-1} b_i f_i + a_{0,d}, \tag{12}$$

where $b_i \in \mathbb{F}_q$ is a linear combination of $a_{i,j}$ which does not involve $f_j$. Hence, $f_d$ is uniquely determined by $f_0, \ldots, f_{d-1}$ as long as $B_0(\xi^d) \neq 0$. Let $S := \{0 \leqslant d \leqslant q - 2 : B_0(\xi^d) = 0\}$. Then it is clear that $|S| \leqslant s - 1$ since the order of $\xi$ is $q - 1$ and $B_0(X)$ has degree at most $s - 1$. Thus, $B_0(\xi^j) \neq 0$ if and only if $j \mod (q-1) \notin S$; and in this case $f_j$ is a fixed affine linear combination of $f_i$ for $0 \leqslant i < j$. Note that $B_0(X)$ has at most $(s-1) \left\lceil \frac{k}{q-1} \right\rceil$ roots among $\{\xi^i : i = 0, 1, \ldots, k-1\}$. It follows that the set of solutions $(f_0, f_1, \ldots, f_{k-1})$ is an affine space $W \subset \mathbb{F}_q^k$, and the dimension of $W$ is at most $(s-1) \left\lceil \frac{k}{q-1} \right\rceil$.

The fact that $W$ is $(s, q-1)$-smooth follows from (12) and noting that the coefficients $b_{d-j}$ for $j \geqslant 1$ in that equation are given by $B_j(\xi^{d-j})$ where $B_j(X) := a_{1,j} + a_{2,j}X + \cdots + a_{s,j}X^{s-1}$. Therefore, once the values of $f_i$, $0 \leqslant i < (j-1)(q-1)$ are fixed, the possible choices for the next block of $(q-1)$ coordinates, $f_{(j-1)(q-1)}, \cdots, f_{j(q-1)-1}$, lie in an affine shift of a fixed subspace of dimension at most $(s-1)$. Further, this shift is an easily computed affine linear combination of the $f_i$'s in the previous blocks. This implies the efficient computability of the claimed representation of $W$.

Finally, by the choice of $D$ in (7), the total number of possible $(A_0, A_1, \ldots, A_s)$ and hence the number of possible functional equations (11), is at most $q^{N(m-s+1)+s+1} \leqslant q^{Nm+s+1}$. Therefore, the number of possible candidate subspaces $W$ is also at most $q^{Nm+s+1}$. □

Combining Lemmas 3.6 and 3.7, we conclude, after some simple calculations, that one can find a representation of the $(s, q-1)$-periodic subspace containing all candidate messages $(f_0, f_1, \ldots, f_{k-1})$ in polynomial time, when the fraction of errors $\tau = 1 - t/N$ satisfies

$$\tau \leqslant \frac{s}{s+1} - \frac{s}{s+1} \frac{k}{N(m-s+1)} - \frac{3m}{m-s+1} \frac{g_e}{mN}. \tag{13}$$

**Pruning the subspace.** Applying Lemma 3.7 directly we would get a list size bound of $\approx q^{sk/q}$ which would be super-polynomial in the code length unless $k = O(q)$. Thus this idea does not directly allow us to get good list decodable codes while keeping the base field size small or achieve a list size that grows polynomially in $s$. Instead what we show is that by only encoding $(f_0, f_1, \ldots, f_{k-1}) \in \mathbb{F}_q^k$ that are restricted to belong to a special *subspace-evasive set*, we can (i) bring down the list size, *and* (ii) find this list efficiently in polynomial time (and further the exponent of the polynomial is independent of $\varepsilon$, the gap to capacity). To this end, we develop the necessary machinery concerning subspace evasive sets next. Later, in Section 5, we combine these subspace evasive sets with our folded Hermitian codes to get good list-decodable codes.

## 4 Subspace evasive sets with additional structure

Let us first recall the notion of "ordinary" subspace-evasive sets from [Gur11].

**Definition 4.** *A subset $S \subset \mathbb{F}_q^k$ is said to be $(d, \ell)$-subspace-evasive if for all d-dimensional affine subspaces $W$ of $\mathbb{F}_q^k$, we have $|S \cap W| \leqslant \ell$.*

We next define the notion of evasiveness w.r.t a collection of subspaces instead of all subspaces of a particular dimension.

**Definition 5.** *Let $\mathcal{F}$ be a family of (affine) subspaces of $\mathbb{F}_q^k$, each of dimension at most $d$. A subset $S \subset \mathbb{F}_q^k$ is said to be $(\mathcal{F}, d, \ell)$-evasive if for all $W \in \mathcal{F}$, we have $|S \cap W| \leqslant \ell$.*

### 4.1 Hierarchical subspace-evasive sets

The key to pruning the list to a small size is the notion of a *hierarchical subspace-evasive set*, which is defined as a subset of $\mathbb{F}_q^k$ with the property that some of its prefixes are subspace-evasive with respect to $(s, \Delta)$-periodic subspaces. We will show how the special subspace-evasive sets help towards pruning the list in our list decoding context in Section 4.5.

**Definition 6.** *Let $\mathcal{F}$ be a family of $(s, \Delta)$-periodic subspaces of $\mathbb{F}_q^k$. A subset $S \subset \mathbb{F}_q^k$ is said to be $(\mathcal{F}, s, \Delta, L)$-h.s.e (for hierarchically subspace evasive for block size $\Delta$) if for every affine subspace $W \in \mathcal{F}$, the following bound holds for $j = 1, 2, \ldots, b$:*

$$|\mathrm{proj}_{j\Delta}(S) \cap \mathrm{proj}_{j\Delta}(W)| \leqslant L \ .$$

### 4.2 Random sets are hierarchically subspace evasive

Our goal is to give a randomized construction of large h.s.e sets that works with high probability, with the further properties that one can index into elements of this set efficiently (necessary for efficient encoding), and one can check membership in the set efficiently (which is important for efficient decoding).

An easy probabilistic argument, see [Gur11], shows that a random subset of $\mathbb{F}_q^k$ of size about $q^{(1-\zeta)k}$ is $(d, O(d/\zeta))$-subspace evasive with high probability. As a warmup, let us work out the similar proof for the case when we have only to avoid a not too large family $\mathcal{F}$ of all possible $d$-dimensional affine subspaces. The advantage is that the guarantee on the intersection size is now $O(1/\zeta)$ and independent of the dimension $d$ of the subspaces one is trying to evade.

**Lemma 4.1.** *Let $\zeta \in (0,1)$ and $k$ be a large enough positive integer. Let $\mathcal{F}$ be a family of affine subspaces of $\mathbb{F}_q^k$, each of dimension at most $d \leqslant \zeta k/2$, with $|\mathcal{F}| \leqslant q^{ck}$ for some positive constant $c$.*

*Let $\mathcal{W}$ be a random subset of $\mathbb{F}_q^k$ chosen by including each $x \in \mathbb{F}_q^k$ in $\mathcal{W}$ with probability $q^{-\zeta k}$. Then with probability at least $1 - q^{-ck}$, $\mathcal{W}$ satisfies both the following conditions: (i) $|\mathcal{W}| \geqslant q^{(1-2\zeta)k}$, and (ii) $\mathcal{W}$ is $(\mathcal{F}, d, 4c/\zeta)$-evasive.*

*Proof.* The first part follows by noting that the expected size of $\mathcal{W}$ equals $q^{(1-\zeta)k}$ and a standard Chernoff bound calculation. For the second part, fix an affine subspace $S \subseteq \mathcal{F}$ of dimension at most $d$, and a subset $T \subseteq S$ of size $t$, for some parameter $t$ to be specified shortly. The probability that $\mathcal{W} \supseteq T$ equals $q^{-\zeta kt}$. By a union bound over the at most $q^{ck}$ choices for the affine subspace $S \in \mathcal{F}$, and the at most $q^{dt}$ choices of $t$-element subsets $T$ of $S$, we get that the probability that $\mathcal{W}$ is not $(\mathcal{F}, d, t)$-evasive is at most $q^{ck+dt} \cdot q^{-\zeta kt} \leqslant q^{ck} q^{-\zeta kt/2}$ since $d \leqslant \zeta k/2$. Choosing $t = \lceil 4c/\zeta \rceil$, this quantity is bounded from above by $q^{-ck}$. □

## 4.3 Pseudorandom construction of large h.s.e subsets

We next turn to the pseudorandom construction of large h.s.e subsets. Suppose, for some fixed subset $\mathcal{F}$ of $(s, \Delta)$-periodic subspaces of $\mathbb{F}_q^k$, we are interested in an $(\mathcal{F}, s, \Delta, \ell)$-h.s.e subset of $\mathbb{F}_q^k$ of size $\approx q^{(1-\zeta)k}$ for a constant $\zeta$, $1/\Delta < \zeta < 1$. For simplicity, let us assume that the block size $\Delta$ divides $k$, though arbitrary $k$ can be easily handled. (We will also ignore floors and ceilings in the description to avoid notational clutter; those are easy to accommodate and do not affect any of the claims.) Define $b = \frac{k}{\Delta}$ to be the number of blocks. The parameters $b, \Delta, k$ and field size $q$ will be considered fixed for the rest of the discussion in this section.

Our construction will use some *arbitrary fixed* subsets $\Lambda_1, \Lambda_2, \ldots, \Lambda_b$ where $\Lambda_i \subseteq \mathbb{F}_{q^{i\Delta}}$ with $|\Lambda_i| = q^{(i-\zeta)\Delta}$. The only requirement from the subsets $\Lambda_i$ is that membership in them can be checked using at most $\text{poly}(i\Delta)$ operations over $\mathbb{F}_q$.

The random part of the construction will consist of two sets of mutually independent, random polynomials $P_1, P_2, \ldots, P_b$ and $Q_1, Q_2, \ldots, Q_b$ where $P_i, Q_i \in \mathbb{F}_{q^{i\Delta}}[T]$ are random polynomials of degree $\lambda$ for $1 \leqslant i \leqslant b$. [3] The degree parameter will be chosen to be $\lambda = \Theta(k)$.

The key fact we will use about the random polynomials $P_i$'s is the following, which follows by virtue of the $\lambda$-wise independence of the values of a random degree $\lambda$ polynomial.

**Fact 4.2.** *For a fixed subset $T \subseteq \mathbb{F}_{q^{i\Delta}}$ with $|T| \leqslant \lambda$, the values $\{P_i(\alpha)\}_{\alpha \in T}$ are independent random values in $\mathbb{F}_{q^{i\Delta}}$.*

In what follows we assume that, for $i = 1, 2, \ldots, b$, some fixed bases of the fields $\mathbb{F}_{q^{i\Delta}}$ have been chosen, giving us some canonical $\mathbb{F}_q$-linear injective maps $\rho_i : \mathbb{F}_q^{i\Delta} \to \mathbb{F}_{q^{i\Delta}}$.

**Definition 7.** *Given the polynomials $P_1, P_2, \ldots, P_b$, define the subset $\Gamma(P_1, P_2, \ldots, P_b)$ by*

$$\{\mathbf{y} = (y_1, y_2, \ldots, y_b) \in \mathbb{F}_q^k \mid y_j \in \mathbb{F}_q^\Delta, \ P_j(\rho_j(y_1 \circ y_2 \circ \cdots \circ y_j)) \in \Lambda_j \ for \ j = 1, 2, \ldots, b\} \ .$$

---

[3] We will assume that representations of the necessary extension fields $\mathbb{F}_{q^{i\Delta}}$ are all available. For this purpose, we only need irreducible polynomials over $\mathbb{F}_q$ of degree $i\Delta$, which can be constructed by picking random polynomials and checking them for irreducibility. Our final construction is anyway randomized, so the randomized nature of this step does not affect the results.

Given the above definition, our final h.s.e set will be defined as follows (we suppress the dependence of $H$ on $P_i, Q_i$ for convenience):

$$H \stackrel{\text{def}}{=} \Gamma(P_1, P_2, \ldots, P_b) \cap \Gamma(Q_1, Q_2, \ldots, Q_b) \ . \tag{14}$$

The reason for defining the set as the intersection of two $\Gamma$ sets will become clear later on. We will later modify the construction slightly to ensure also the efficient encoding property that we seek (but it is cleaner to first present the construction without this extra concern).

We first note that it is highly likely that the set $H$ is large, and then establish the h.s.e property.

**Lemma 4.3.** *With probability at least $1 - q^{-\Omega(k)}$ over the choice of $\{P_i, Q_i\}_{1 \leqslant i \leqslant b}$, we have*

$$|H| \geqslant q^{(1-3\zeta)k} \ .$$

*Proof.* For each $\mathbf{y} \in \mathbb{F}_q^k$, define the indicator random variable $I_\mathbf{y}$ for the event that $\mathbf{y} \in \Gamma$. We have $\mathbb{E}[I_\mathbf{y}] = (q^{-\zeta\Delta})^{2b} = q^{-2\zeta k}$. Therefore, $\mathbb{E}[|H|] = \sum_{\mathbf{y} \in \mathbb{F}_q^k} \mathbb{E}[I_\mathbf{y}] = q^{(1-2\zeta)k}$. For degree $\lambda \geqslant 2$, the random variables $I_\mathbf{y}$'s are pairwise independent, so by Chebyshev's inequality, we have that $|\Gamma| \geqslant q^{(1-3\zeta)k}$ with probability at least $1 - q^{-\Omega(k)}$. □

We now move on to the main claim about the h.s.e property of our construction.[4]

**Theorem 4.4.** *Let $\zeta \in (0, 1)$ and $s$ be a positive integer satisfying $s < \zeta\Delta/10$. Let $\mathcal{F}$ be a subset of at most $q^{ck}$ $(s, \Delta)$-periodic subspaces of $\mathbb{F}_q^k$ for some positive constant $c$. Suppose that the parameters satisfy the condition $q^{\zeta\Delta} \geqslant (2qck)^{10/9}$. Then with probability $1 - \exp(-\Omega(k))$ over the choice of random polynomials $\{P_i, Q_i\}_{1 \leqslant i \leqslant b}$ each of degree $\lambda > ck$, the set $H$ defined in (14) is $(\mathcal{F}, s, \Delta, L)$-h.s.e and $(\mathcal{F}, sb, \ell)$-evasive for $L \leqslant ck$ and $\ell \leqslant 20c/\zeta$.*

*Proof.* We will prove that w.h.p over the choice of $P_i$'s, the subset $\Gamma \stackrel{\text{def}}{=} \Gamma(P_1, P_2, \ldots, P_b)$ is $(\mathcal{F}, s, \Delta, L)$-h.s.e for $L = ck$, and this will imply the same for $H$ as $H \subseteq \Gamma(P_1, P_2, \ldots, P_b)$. We will then prove that conditioned on $\Gamma$ being $(\mathcal{F}, s, \Delta, L)$-h.s.e, with high probability over the choice of the $Q_i$'s, $H$ will intersect any subspace in $\mathcal{F}$ at less than $O(1/\zeta)$ points. (Note that every subspace in $\mathcal{F}$ has dimension at most $sb$ by Observation 2.1.) Together, these steps will imply the claim of the theorem.

For the first step, it suffices to show that w.h.p, the following holds: For every $(s, \Delta)$-smooth subspace $W \subseteq \mathbb{F}_q^k$ that belongs to $\mathcal{F}$, we have

$$|\Gamma_i \cap W_i| \leqslant L \text{ for } i = 1, 2, \ldots, b$$

where $W_i = \text{proj}_{i\Delta}(W)$ and $\Gamma_i = \text{proj}_{i\Delta}(\Gamma)$. (Recalling the definition of $\Gamma$, this means that

$$\Gamma_i = \{\mathbf{z} \in \mathbb{F}_q^{i\Delta} \mid P_i(\rho_i(\mathbf{z})) \in \Lambda_i \text{ and } \text{proj}_{j\Delta}(\mathbf{z}) \in \Gamma_j \text{ for } 1 \leqslant j < i\} \ .)$$

We will establish this by induction. For the base case $i = 1$, this is just the standard argument using the $\lambda$-wise independence of the set $\Gamma_i$. By the choice of the degree $\lambda$ we made, $L < \lambda$. For each fixed set of $L$ elements, the events that they all belong to $\Gamma_1$ are independent and each occurs with probability equal to the density of $\Lambda_1 \subseteq \mathbb{F}_{q^\Delta}$ which $q^{-\zeta\Delta}$. Therefore, for a fixed subspace $U \subseteq \mathbb{F}_q^\Delta$ of dimension $s$, the probability that $|U \cap \Gamma_1| \geq L$ is at most

$$\binom{q^s}{L} q^{-\zeta\Delta L} \leq q^{sL} q^{-\zeta\Delta L} \leq q^{-9sL} ,$$

using $s \leq \zeta\Delta/10$. Since there are at most $|\mathcal{F}|$ candidates $U$ which are of the form $\text{proj}_\Delta(W)$ for $W \in \mathcal{F}$, by a union bound over such $U$, we conclude that the probability that $|\Gamma_1 \cap W_1| \geq L$ for some $W \in \mathcal{F}$ is at most $q^{ck} q^{-9sL} \leq q^{-\Omega(k)}$ for the choice $L = ck$.

---

[4] We have not attempted to optimize the constants in the conditions stated in the theorem.

Suppose now that $i > 1$ and we condition on the fact that for every $W \in \mathcal{F}$, we have $|\Gamma_{i-1} \cap W_{i-1}| \leq L$. Let us first fix $W \in \mathcal{F}$ and upper bound the probability that $|\Gamma_i \cap W_i| > L$. Since $|\Gamma_{i-1} \cap W_{i-1}| \leq L$, the number of candidate elements in $\Gamma_i \cap W_i$ is at most $L \cdot q^s$, since there are at most $L$ possibilities for the first $(i-1)\Delta$ coordinates, and by the definition of $(s,\Delta)$-smoothness (Definition 1), for each of these there are at most $q^s$ possibilities for the last $\Delta$ coordinates. The probability that some $L$ of these candidates actually belong to $\Gamma_i$, and thus the probability that $|\Gamma_i \cap U_i| \geq L$, is at most $(Lq^s)^L \cdot q^{-\zeta\Delta L}$. Taking a union bound over all $W \in \mathcal{F}$, we have $|\Gamma_i \cap W_i| \leq L$ for all $W \in \mathcal{F}$ except with probability at most

$$q^{ck}(L \cdot q^{(s-\zeta\Delta)})^L \leq q^{ck}(L \cdot q^{-0.9\zeta\Delta})^L = (q \cdot ck \cdot q^{-0.9\zeta\Delta})^{ck} \leq 2^{-ck}$$

where the last step used the hypothesis that $q^{\zeta\Delta} \geq (2cqk)^{10/9}$, and the previous step used that $L = ck$. This finishes the proof that $\Gamma$ is $(\mathcal{F}, s, \Delta, L)$-h.s.e except with $2^{-\Omega(k)}$ probability.

Suppose we now condition on the event that $|\Gamma \cap W| \leq L$ (which we showed happens with high probability) after the $P_i$'s are picked. Let us now prove that w.h.p over the choice of the $Q_i$'s, $H = \Gamma \cap \Gamma(Q_1, Q_2, \ldots, Q_b)$ intersects every $W \in \mathcal{F}$ at not more than $\ell$ points. Note that $H \cap W \subseteq \Gamma \cap W$, so only the (at most $L$) elements of $\Gamma \cap W$ can belong to $H \cap W$. Fixing a $W \in \mathcal{F}$, the probability that at least $\ell$ elements of $\Gamma \cap W$ belong to $H$ is at most $\binom{L}{\ell} \cdot q^{-\zeta k \ell}$ since the probability that a fixed $\mathbf{y} \in \mathbb{F}_q^k$ belongs to $\Gamma(Q_1, \ldots, Q_b)$ equals $(q^{-\zeta\Delta})^b = q^{-\zeta k}$. By a union bound over all $W \in \mathcal{F}$, the conditional probability that $|H \cap W| > \ell$ for some $W \in \mathcal{F}$ is at most

$$q^{ck} L^\ell q^{-\zeta k \ell} \leq q^{ck} q^{0.9\zeta\Delta \ell} q^{-\zeta k \ell} \leq q^{ck} q^{-0.1\zeta k \ell} .$$

Choosing $\ell = 20c/\zeta$, this probability is at most $q^{-ck}$. $\square$

## 4.4 Efficient encoding of h.s.e. subsets

The construction of the h.s.e set in (14) allows for efficient membership checks in $H$ — once the $P_i$'s and $Q_i$'s are sampled, it follows from Definition 7 that one can check membership in $\Gamma(P_1, P_2, \ldots, P_b)$ and $\Gamma(Q_1, Q_2, \ldots, Q_b)$. The construction does not, however, provide an efficient method to index into elements of $H$, which is necessary for efficient encoding of messages into elements of $H$. In this section we will show that w.h.p. $H$ contains a certain subset that permits efficient encoding.

We will describe this subset of $H$ by giving an encoding map from strings in $\mathbb{F}_q^{(1-3\zeta)k}$ to the set. We will then prove that the map is well-defined.

**Definition 8** (Encoding into the h.s.e set). *Given the polynomials $P_1, P_2, \ldots, P_b$ and $Q_1, \ldots, Q_b$, and the subsets $\Lambda_i \subseteq \mathbb{F}_{q^{i\Delta}}$, the encoding of $\mathbf{x} = (x_1, x_2, \ldots, x_b)$ where $x_i \in \mathbb{F}_q^{(1-3\zeta)\Delta}$, proceeds as follows:*

*For $i = 1, 2, \ldots, b$*

- *Let $\beta_i \in \mathbb{F}_q^{3\zeta\Delta}$ be the lexicographically first string such that $P_i(\rho_i(x_1 \circ \beta_1 \circ \cdots \circ x_i \circ \beta_i)) \in \Lambda_i$ and $Q_i(\rho_i(x_1 \circ \beta_1 \circ \cdots \circ x_i \circ \beta_i)) \in \Lambda_i$. If no such $\beta_i$ exists, fail.*

*Output $x_1 \circ \beta_1 \circ x_2 \circ \beta_2 \circ \cdots \circ x_b \circ \beta_b \in \mathbb{F}_q^k$ as the encoding of $\mathbf{x}$.*

*We will denote the above encoding map by HSE and refer to $\Delta$ as its period size (we suppress the dependence of the map on the $P_i$'s and $Q_i$'s for convenience).*

We now prove that the above map is well-defined, in the sense that the required $\beta_i$'s will exist with high probability. Note that HSE, when it is well-defined, is an injective map.

**Lemma 4.5.** *Suppose the parameters satisfy $q^{\zeta\Delta} \geqslant 3q\lambda$ and $\lambda \geqslant 6k$. For every choice of the subsets $\Lambda_i \subseteq \mathbb{F}_{q^{i\Delta}}$ with $|\Lambda_i| = q^{(i-\zeta)\Delta}$, the following holds with probability at least $1 - q^{-\Omega(k)}$ over the choice of the random degree $\lambda$ polynomials $\{P_i : 1 \leqslant i \leqslant b\}$ and $\{Q_i : 1 \leqslant i \leqslant b\}$: For all $\mathbf{x} = (x_1, x_2, \ldots, x_b)$ where $x_i \in \mathbb{F}_q^{(1-3\zeta)\Delta}$, the above procedure successfully encodes $\mathbf{x}$.*

*The encoding complexity is at most $O(q^{3\zeta\Delta} \lambda k^2 \log^2 k)$ operations over $\mathbb{F}_q$.*

*Proof.* First, let us note that the encoding complexity is as claimed when the encoding succeeds. Given $\mathbf{a} \in \mathbb{F}_{q^{i\Delta}}$ we can compute $P_i(\mathbf{a})$ and $Q_i(\mathbf{a})$ using at most $O(\lambda k \log^2 k)$ $\mathbb{F}_q$-operations. We can pick $\Lambda_i$ so that membership of an element of $\mathbb{F}_{q^{i\Delta}}$ in $\Lambda_i$ can be checked using $O(k^2)$ operations. Therefore, for each $i = 1, 2, \ldots, b$, the search for $\beta_i$ takes $q^{3\zeta\Delta} \cdot O(\lambda k \log^2 k)$ operations over $\mathbb{F}_q$. This gives a bound of $O(q^{3\zeta\Delta} \lambda k^2 \log^2 k)$ operations over $\mathbb{F}_q$ for the total encoding complexity.

Let us now prove that $\mathsf{HSE}(\mathbf{x})$ exists for all $\mathbf{x}$ with high probability, taken over the choice of the random polynomials $P_i$, $Q_i$. Fix an $\mathbf{x} \in \mathbb{F}_q^{(1-2\zeta)k}$. For $1 \leqslant j \leqslant b$, conditioned on the choice of $\beta_1, \beta_2, \ldots, \beta_{j-1}$, the probability that a fixed $\alpha \in \mathbb{F}_q^{3\zeta\Delta}$ satisfies $P_j(\rho_j(x_1 \circ \beta_1 \circ \cdots x_{j-1} \circ \beta_{j-1} \circ x_j \circ \alpha)) \in \Lambda_j$ and $Q_j(\rho_j(x_1 \circ \beta_1 \circ \cdots x_{j-1} \circ \beta_{j-1} \circ x_j \circ \alpha)) \in \Lambda_j$ is $q^{-2\zeta\Delta}$. Let $N_j$ be the random variable equal to the number of elements $\alpha \in \mathbb{F}_q^{3\zeta\Delta}$ such that $P_j(\rho_j(x_1 \circ \beta_1 \circ \cdots x_{j-1} \circ \beta_{j-1} \circ x_j \circ \alpha)) \in \Lambda_j$ and $Q_j(\rho_j(x_1 \circ \beta_1 \circ \cdots x_{j-1} \circ \beta_{j-1} \circ x_j \circ \alpha)) \in \Lambda_j$. The expected value of $N_j$ equals $\mu \stackrel{\text{def}}{=} q^{\zeta\Delta}$. Note that $\mu \geqslant 3q\lambda$ by the hypothesis in the lemma.

By concentration inequalities for $\lambda$-wise independent random variables, see for example [BR94, Lemma 2.3], the probability that $N_j = 0$ is at most

$$8 \cdot \left(\frac{\lambda\mu + \lambda^2}{\mu^2}\right)^{\lambda/2} \leqslant 8 \cdot \left(\frac{1}{2q}\right)^{\lambda/2} \leqslant q^{-3k} .$$

Summing up these conditional probabilities for $j = 1, 2, \ldots, b$, the probability that $\mathsf{HSE}(\mathbf{x})$ does not exist is at most $b \cdot q^{-3k} \leqslant q^{-2k}$. Finally, a union bound over all $\mathbf{x} \in \mathbb{F}_q^{(1-3\zeta)k}$ shows that the probability that *some* $\mathbf{x}$ does not have a valid encoding $\mathsf{HSE}(\mathbf{x})$ is at most $q^{-k}$. □

Suppose the polynomials $P_i, Q_i$, $1 \leqslant i \leqslant b$, are such that $\mathsf{HSE}(\mathbf{x})$ is defined for all $\mathbf{x} \in \mathbb{F}_q^{(1-3\zeta)k}$. This implies that the set $\Theta = \{\mathsf{HSE}(\mathbf{x}) \mid \mathbf{x} \in \mathbb{F}_q^{(1-3\zeta)k}\}$ has size $q^{(1-3\zeta)k}$. Also note that $\Theta \subseteq H = \Gamma(P_1, P_2, \ldots, P_b) \cap \Gamma(Q_1, \ldots, Q_b)$. Thus if $H$ is $(\mathcal{F}, s, \Delta, L)$-h.s.e and $(\mathcal{F}, sb, \ell)$-evasive then so is $\Theta$. Therefore the claim of Theorem 4.4 also holds for $\Theta$. Putting these together, we have the following main result on the pseudorandom construction of efficiently encodable h.s.e sets. The elements of this h.s.e set will give the subset of messages that we will encoded by the folded algebraic codes for our final list-decodable code construction. The proof is an immediate consequence of Theorem 4.4 and Lemma 4.5 (note that the stated conditions (15) on the parameters meet the requirements of both Theorem 4.4 and Lemma 4.5).

**Theorem 4.6** (Main construction of h.s.e. subsets). *Suppose $b, c, \Delta, k, s$ are positive integers and $\zeta \in (0,1)$ such that following requirements are met:*

$$k = b\Delta; \quad s < \zeta\Delta/10; \quad q^{\zeta\Delta} \geqslant (2cqk)^{10/9} \;. \tag{15}$$

*Let $\mathcal{F}$ be a family of $(s, \Delta)$-periodic subspaces of $\mathbb{F}_q^k$ with $|\mathcal{F}| \leqslant q^{ck}$. Then, for a random and independent choice of polynomials $P_i, Q_i \in \mathbb{F}_{q^{i\Delta}}[T]$ of degree $\lambda = \max\{6k, ck+1\}$ and any subsets $\Lambda_i$ of size $q^{(i-\zeta)\Delta}$ for $i = 1, 2, \ldots, b$, the following conditions both hold with probability at least $1 - 2^{-\Omega(k)}$:*

1. $\mathsf{HSE}: \mathbb{F}_q^{(1-3\zeta)k} \to \mathbb{F}_q^k$ *from Definition 8 is a well-defined injective map, and can be computed using $O(q^{3\zeta\Delta} k^3 \log^2 k)$ operations over $\mathbb{F}_q$.*

2. *The set $H = \Gamma(P_1, \ldots, P_b) \cap \Gamma(Q_1, \ldots, Q_b)$, and in particular the image of $\mathsf{HSE}$, is a $(\mathcal{F}, s, \Delta, ck)$-h.s.e and a $(\mathcal{F}, sb, 20c/\zeta)$-evasive subset of $\mathbb{F}_q^k$.*

### 4.5 Efficient computation of intersection with h.s.e. subsets

The key aspect which makes h.s.e subsets useful in our context to prune the affine space of candidate messages, and indeed motivated the exact specifics of the definition and construction, is the following claim which shows that intersection of a $(s, \Delta)$-periodic subspace with our h.s.e set can found efficiently.

**Lemma 4.7.** *Suppose polynomials $P_i, Q_i$, $i = 1, 2, \ldots, b$, of degree $\lambda = \max\{6k, ck+1\}$ have been picked so that the map $\mathsf{HSE}$ satisfies the conditions of Theorem 4.6 w.r.t some family $\mathcal{F}$ of at most $q^{ck}$ affine subspaces of $\mathbb{F}_q^k$ each of which is $(s, \Delta)$-periodic. Then given a representation of $W \in \mathcal{F}$ (as in Definition 1), we can find the list of at most $O(c/\zeta)$ values of $\mathbf{x} \in \mathbb{F}_q^{(1-3\zeta)k}$ such that $\mathsf{HSE}(\mathbf{x}) \in \mathcal{F}$ using $O\left(c^2(kq^s + q^{3\zeta\Delta})k^3 \log^2 k\right)$ operations over $\mathbb{F}_q$.*

*Proof.* The fact that there are at most $\ell$ solutions $\mathbf{x}$ follows immediately from the fact that the image of $\mathsf{HSE}$ is $(\mathcal{F}, sb, \ell)$-evasive. So we only need to argue about the time complexity.

For $1 \leqslant i \leqslant b$, define $H_i = \mathrm{proj}_{i\Delta}(H)$ where $H = \Gamma(P_1, \ldots, P_b) \cap \Gamma(Q_1, \ldots, Q_b)$. Likewise, define $W_i = \mathrm{proj}_{i\Delta}(W)$. To compute the intersection $H \cap W$ list efficiently, we iteratively find $W_i \cap H_i$ for $i = 1, 2, \ldots, b$ as follows. Recall that we know that $|H_i \cap W_i| \leqslant L$ for each $i$ as $H$ is $(\mathcal{F}, s, \Delta, L)$-h.s.e. For each of the at most $L = ck$ candidates in $W_{i-1} \cap H_{i-1}$, as $W$ is $(s, \Delta)$-periodic, there are at most $q^s$ possible extensions to the next block of $(q-1)$ coordinates which we can find and list using $O(q^s \cdot k\Delta)$ operations. (The $k\Delta$ term comes from computing the affine shift for the $i$'th block for that particular prefix of $(i-1)\Delta$ symbols.)

We then test each of $L \cdot q^s$ candidates for membership in $\Gamma_i$ which can be done using $O(ck^2 \log^2 k)$ $\mathbb{F}_q$-operations time by evaluating the degree $\lambda$ polynomial and checking that the resulting value belongs to $\Lambda_i$. By the $(\mathcal{F}, s, \Delta, L)$-h.s.e property of $H$ there are at most $L$ of these candidates that can belong to $H_i$, thus bringing our list size back to $L$. The runtime for each iterative step is $O(Lq^s k\Delta + Lq^s ck^2 \log^2 k) = O(cLq^s k^2 \log^2 k)$ $\mathbb{F}_q$-operations, leading to an overall runtime for all $b < k$ stages of $O(cLq^s k^3 \log^2 k)$ operations over $\mathbb{F}_q$ to recover the intersection $H \cap W$. Finally for each $\mathbf{y} \in H \cap W$, we can check if it is the range of HSE by writing $\mathbf{y} = x_1 \circ \beta_1 \circ \cdots \circ x_b \circ \beta_b$ and checking that $\mathsf{HSE}(x_1 \circ x_2 \circ \cdots \circ x_b) = \mathbf{y}$, which takes $O(q^{3\zeta\Delta} ck^3 \log^2 k)$ operations over $\mathbb{F}_q$. □

## 5 Combining folded Hermitian codes and h.s.e sets

Instead of encoding arbitrary $\mathbf{f} \in \mathbb{F}_q^k$ by the folded Hermitian code (Definition 3), we can restrict the messages $\mathbf{f}$ to belong to the range of our h.s.e set, so that the affine space of solutions guaranteed by Lemma 3.7 can be efficiently pruned to a small list. The formal claim is below.

**Theorem 5.1.** *Let $e \geqslant 2$ be an integer, $r \geqslant 2e$ be a large enough prime power, $q = r^2$, and $\zeta \in (1/q, 1)$. Let $k \leqslant q^{\zeta q/2}$ be a positive integer. Let $s, m$ be positive integers satisfying $1 \leqslant s \leqslant m \leqslant q-1$ and $s < \zeta q/12$. Finally let $N$ be an integer satisfying $k + 2er^e \leqslant Nm \leqslant (q-1)r^e$.*

*Consider the code $C_1$ with encoding $E_1 : \mathbb{F}_q^{(1-3\zeta)k} \to (\mathbb{F}_q^m)^N$ defined as*

$$E_1(\mathbf{x}) = \mathsf{FH}(N, k, q, e, m)(\mathsf{HSE}(\mathbf{x})) \ ,$$

*for $\mathsf{HSE} : \mathbb{F}_q^{(1-3\zeta)k} \to \mathbb{F}_q^k$ from Definition 8 for a period size $\Delta = q - 1$.*

*Then, with high probability over the choice of $\mathsf{HSE}$, this code has rate $R = (1-3\zeta)k/(Nm)$, can be encoded in $\mathrm{poly}(Nmq^{\zeta q})$ time, and is $(\tau, \ell)$-list decodable in time $\mathrm{poly}(Nmq^{\zeta q})$ for $\ell \leqslant O(1/(R\zeta))$ and*

$$\tau = \frac{s}{s+1}\left(1 - \frac{k}{N(m-s+1)}\right) - \frac{3m}{m-s+1}\frac{er^e}{mN} \ .$$

*Proof.* This follows by just combining the ingredients we have developed so far. Since $g_e \leqslant er^e$ by (2), the condition on $N, m$ meets the requirement for the construction of the folded Hermitian tower based code in Definition 2.

Whp, the map HSE is well-defined and injective, and so $E_1$ is an injective encoding. The rate of the code is therefore clearly as claimed. With $\Delta = q - 1$, one can check that the conditions of Theorem 4.6 are met for our choice of $\zeta, s, q, k$. By Theorem 4.6 , Part 1, HSE can be computed in time $\mathrm{poly}(Nmq^{\zeta q})$ and hence so can $E_1$ (as FH is efficiently encodable as well).

The claimed value of the error fraction $\tau$ satisfies (13) since the genus is at most $er^e$ by (2). By Lemma 3.7, we know that the candidate messages found by the decoder lie in one of at most $q^{2Nm}$ possible $(s, q-1)$-periodic subspaces. Appealing to Theorem 4.6 and Lemma 4.7 with the choice $c = 2Nm/k = O(1/R)$, we conclude that there is a decoding algorithm running in time $\mathrm{poly}(Nmq^{\zeta q})$ to list decode $C_1$ from a fraction $\tau$ of errors, outputting at most $O(1/(R\zeta))$ messages in the worst-case. □

Let $\varepsilon > 0$ be a small positive constant, and a family of codes of length $N$ (assumed large enough) and rate $R \in (0, 1)$ is sought. Pick $n$ to be a growing parameter.

By picking $s = \Theta(1/\varepsilon)$, $m = \Theta(1/\varepsilon^2)$, $r = \lfloor \log n \rfloor$, $e = \lceil \frac{\log n}{\log \log n} \rceil$, $\zeta = (\log n \log \log n)^{-1}$, $N = \lfloor \frac{(r^2-1)r^e}{m} \rfloor$, and $k$ proportional to $Nm$ in Theorem 5.1, we can conclude the following.

**Corollary 5.2.** *For any $R \in (0, 1)$ and positive constant $\varepsilon \in (0, 1)$, there is a Monte Carlo construction of a family of codes of rate at least $R$ over an alphabet size $(\log N)^{O(1/\varepsilon^2)}$ that are encodable and $(1 - R - \varepsilon, O(R^{-1} \log N \log \log N))$-list decodable in $\mathrm{poly}(N, 1/\varepsilon)$ time, where $N$ is the block length of the code.*

Our promised main result (Theorem 1.1) achieves better parameters than the above — an alphabet size of $\exp(\tilde{O}(1/\varepsilon^2))$ and list-size of $O(1/(R\varepsilon))$. This is based on the Garcia-Stichtenoth tower and is described next.

# 6 Folded codes from the Garcia-Stichtenoth tower

Compared with the Hermitian tower of function fields, the Garcia-Stichtenoth tower of function fields yields folded codes with better parameters due to the fact that the Garcia-Stichtenoth tower is an optimal one in the sense that the ratio of number of rational places against genus achieves the maximal possible value. The construction of folded codes from the Garcia-Stichtenoth tower is almost identical to the one from the Hermitian tower except for one major difference: the redefined code from the Garcia-Stichtenoth tower is constructed in terms of the local expansion at point $P_\infty$, while in the Hermitian case local expansion at $P_0$ is considered. For convenience of the reader, we give a parallel description of folded codes from the Garcia-Stichtenoth tower, while only sketching the identical parts.

## 6.1 Background on Garcia-Stichtenoth tower

Again let $r$ be a prime power and let $q = r^2$. We denote by $\mathbb{F}_q$ the finite field with $q$ elements. The Garcia-Stichtenoth towers that we are going to use for our code construction were discussed in [GS95, GS96]. The reader may refer to [GS95, GS96] for the detailed background on the Garcia-Stichtenoth function tower. There are two optimal Garcia-Stichtenoth towers that are equivalent. For simplicity, we introduce the tower defined by the following recursive equations [GS96]

$$x_{i+1}^r + x_{i+1} = \frac{x_i^r}{x_i^{r-1} + 1}, \quad i = 1, 2, \ldots, e - 1.$$

Put $K_e = \mathbb{F}_q(x_1, x_2, \ldots, x_e)$ for $e \geqslant 2$.

**Rational places.** The function field $K_e$ has at least $r^{e-1}(r^2 - r) + 1$ rational places. One of these is the "point at infinity" which is the unique pole $P_\infty$ of $x_1$ (and is fully ramified). The other $r^{e-1}(r^2 - r)$ come from the rational places lying over the unique zero of $x_1 - \alpha$ for each $\alpha \in \mathbb{F}_q$ with $\alpha^r + \alpha \neq 0$. Note that for every $\alpha \in \mathbb{F}_q$ with $\alpha^r + \alpha \neq 0$, the unique zero of $x_1 - \alpha$ splits completely in $K_e$, i.e., there are $r^{e-1}$ rational places lying over the zero of $x_1 - \alpha$. Let $\mathbb{P}$ be the set of all the rational places lying over the zero of $x_1 - \alpha$ for all $\alpha \in \mathbb{F}_q$ with $\alpha^r + \alpha \neq 0$. Then, intuitively, one can think of the $r^{e-1}(r^2 - r)$ rational places in $\mathbb{P}$ as being given by $e$-tuples $(\alpha_1, \alpha_2, \ldots, \alpha_e) \in \mathbb{F}_q^e$ that satisfy $\alpha_{i+1}^r + \alpha_{i+1} = \frac{\alpha_i^r}{\alpha_i^{r-1}+1}$ for $i = 1, 2, \ldots, e-1$ and $\alpha_1^r + \alpha_1 \neq 0$. For each value of $\alpha \in \mathbb{F}_q$, there are precisely $r$ solutions to $\beta \in \mathbb{F}_q$ satisfying $\beta^r + \beta = \frac{\alpha^r}{\alpha^{r-1}+1}$, so the number of such $e$-tuples is $r^{e-1}(r^2 - r)$ ($r^2 - r$ choices for $\alpha_1$, and then $r$ choices for each successive $\alpha_i$, $2 \leq i \leq e$).

**Riemann-Roch spaces.** As shown in [SAK$^+$01], every function of $K_e$ with a pole only at $P_\infty$ has an expression of the form

$$x_1^a \left( \sum_{i_1=0}^{(e-2)r+1} \sum_{i_2=0}^{r-1} \cdots \sum_{i_e=0}^{r-1} c_{\mathbf{i}} h_1 \frac{x_1^{i_1} x_2^{i_2} \cdots x_e^{i_e}}{\pi_2 \ldots \pi_{e-1}} \right), \tag{16}$$

where $a \geq 0$, $c_{\mathbf{i}} \in \mathbb{F}_q$, and for $1 \leq j < e$, $h_j = x_j^{r-1} + 1$ and $\pi_j = h_1 h_2 \ldots h_j$. Moreover, Shum et al. [SAK$^+$01] present an algorithm running in time polynomial in $l$ that outputs a basis of over $\mathbb{F}_q$ of $\mathcal{L}(lP_\infty)$ explicitly in the above form.

We stress that evaluating elements of $\mathcal{L}(lP_\infty)$ at the rational places of $\mathbb{P}$ is easy: we simply have to evaluate a linear combination of the monomials allowed in (16) at the tuples $(\alpha_1, \alpha_2, \ldots, \alpha_e) \in \mathbb{F}_q^e$ mentioned above. In other words, it is just evaluating an $e$-variate polynomial at a specific subset of $r^{e-1}(r^2 - r)$ points of $\mathbb{F}_q^e$, and can be accomplished in polynomial time.

**Genus.** The genus $g_e$ of the function field $K_e$ is given by

$$g_e = \begin{cases} (r^{e/2} - 1)^2 & \text{if } e \text{ is even} \\ (r^{(e-1)/2} - 1)(r^{(e+1)/2} - 1) & \text{if } e \text{ is odd.} \end{cases}$$

Thus the genus $g_e$ is at most $r^e$. (Compare this with the $er^e$ bound for the Hermitian tower; this smaller genus is what allows to pick $e$ as large as we want in the Garcia-Stichtenoth tower, while keeping the field size $q$ fixed.)

**A useful automorphism.** Let $\gamma$ be a primitive element of $\mathbb{F}_r$ and consider the automorphism $\sigma \in \text{Aut}(K_e/\mathbb{F}_q)$ defined by

$$\sigma : x_i \mapsto \gamma^{(r+1)r^{i-1}} x_i \quad \text{for } i = 1, 2, \ldots, e.$$

Then the order of $\sigma$ is $r - 1$ and furthermore, we have the following facts:

(i) $\sigma$ keeps $P_\infty$ unchanged, i.e., $P_\infty^\sigma = P_\infty$;

(ii) Let $\mathbb{P}$ be the set of all the rational places lying over $x_1 - \alpha$ for all $\alpha \in \mathbb{F}_q$ with $\alpha^r + \alpha \neq 0$. Then $|\mathbb{P}| = (r-1)r^e$. Moreover, $\sigma$ divides $\mathbb{P}$ into $r^e$ orbits and each orbit has $r - 1$ places. For an integer $m$ with $1 \leq m \leq r - 1$, we can label $Nm$ distinct elements

$$P_1, P_1^\sigma, \ldots, P_1^{\sigma^{m-1}}, \ldots, P_N, P_N^\sigma, \ldots, P_N^{\sigma^{m-1}}$$

in $\mathbb{P}$, as long as $N \leq r^e \lfloor \frac{r-1}{m} \rfloor$.

The folded codes from the Garcia-Stichtenoth tower are defined similarly to the Hermitian case.

**Definition 9** (Folded codes from the Garcia-Stichtenoth tower). *Assume that $m, k, N$ are positive integers satisfying $1 \leqslant m \leqslant r - 1$ and $l/m < N \leqslant r^e \lfloor \frac{r-1}{m} \rfloor$. The folded code from $K_e$ with parameters $N, l, q, e, m$, denoted by $\widetilde{\mathsf{FGS}}(N, l, q, e, m)$, encodes a message function $f \in \mathcal{L}(lP_\infty)$ as*

$$f \mapsto \left( \begin{bmatrix} f(P_1) \\ f(P_1^\sigma) \\ \vdots \\ f(P_1^{\sigma^{m-1}}) \end{bmatrix}, \begin{bmatrix} f(P_2) \\ f(P_2^\sigma) \\ \vdots \\ f(P_2^{\sigma^{m-1}}) \end{bmatrix}, \ldots, \begin{bmatrix} f(P_N) \\ f(P_N^\sigma) \\ \vdots \\ f(P_N^{\sigma^{m-1}}) \end{bmatrix} \right) \in \left( \mathbb{F}_q^m \right)^N . \tag{17}$$

Then we have a similar result on parameters of $\widetilde{\mathsf{FGS}}(N, l, q, e, m)$.

**Lemma 6.1.** *The above code $\widetilde{\mathsf{FGS}}(N, l, q, e, m)$ is an $\mathbb{F}_q$-linear code over alphabet size $q^m$, rate at least $\frac{l - g_e + 1}{Nm}$, and minimum distance at least $N - \frac{l}{m}$.*

## 6.2 Redefining the code in terms of local expansion at $P_\infty$

In the Hermitain case, we use coefficients of its power series expansion around $P_0$. However, for the Garcia-Stictenoth tower we do not have such a nice point $P_0$. Fortunately, we can use point $P_\infty$ to achieve our mission.

Again for our decoding, we will actually recover the message $f \in \mathcal{L}(lP_\infty)$ in terms of the coefficients of its power series expansion around $P_\infty$

$$f = T^{-l}(f_0 + f_1 T + f_2 T^2 + \cdots)$$

where $T := \frac{1}{x_e}$ is the local parameter at $P_\infty$ (which means that $x_e$ has exactly one pole at $P_\infty$, i.e., $\nu_{P_\infty}(x_e) = -1$).

In this case we can also show that one can efficiently move back-and-forth between the representation of $f \in \mathcal{L}(lP_\infty)$ in terms of a basis for $\mathcal{L}(lP_\infty)$ and its power series representation $(f_0, f_1, \ldots)$ around $P_\infty$. Since the mapping $f \mapsto (f_0, f_1, \ldots)$ is $\mathbb{F}_q$-linear, it suffices to compute the local expansion at $P_\infty$ of a basis for $\mathcal{L}(lP_\infty)$.

**Lemma 6.2.** *For any $n$, one can compute the first $n$ terms of the local expansion of the basis elements (16) at $P_\infty$ using $\mathrm{poly}(n)$ operations over $\mathbb{F}_q$.*

*Proof.* First let $h$ be a nonzero function in $\mathbb{F}_q(x_1, x_2, \ldots, x_e)$ with $\nu_{P_\infty}(h) = v \in \mathbb{Z}$. Assume that the local expansion $h = T^v \sum_{j=0}^\infty a_j T^j$ is known. To find the local expansion $\frac{1}{h} = T^{-v} \sum_{j=0}^\infty c_j T^j$. Consider the identity

$$1 = \left( \sum_{j=0}^\infty c_j T^j \right) \left( \sum_{j=0}^\infty a_j T^j \right).$$

Then by comparing the coefficients of $T^i$ in the above identity, one has $c_0 = a_0^{-1}$ and $c_i = -a_0^{-1}(c_{i-1} a_1 + \cdots + c_0 a_i)$ can be easily computed recursively for all $i \geqslant 1$.

Thus, by the structure of the basis functions in (16), it is sufficient to find an algorithm of efficiently finding local expansions of $x_i$ at $P_\infty$ for every $i = 1, 2, \ldots, e$. We can inductively find the local expansions of $x_i$ at $P_\infty$ as follows. We note that $\nu_{P_\infty}(x_i) = -r^{e-i}$ for $i = 1, 2, \ldots, e$.

For $i = e$, $x_e$ has the local expansion $\frac{1}{T}$ at $P_\infty$.

Now assume that we know the local expansion of $x_i$. Then we can easily compute the local expansion of $x_i^r + x_i$ and hence the local expansion of $1/(x_i^r + x_i)$. Let us assume that $1/(x_i^r + x_i)$ has local expansion $1/(x_i^r + x_i) = T^{r^{e-i+1}} \sum_{j=0}^{\infty} \alpha_j T^j$ at $P_\infty$ for some $\alpha_i \in \mathbb{F}_q$. Assume that $1/x_{i-1}$ has the local expansion $1/x_{i-1} = T^{r^{e-i+1}} \sum_{j=0}^{\infty} \beta_j T^j$. To find $\beta_j$, we consider the identity

$$T^{r^{e-i+1}} \sum_{j=0}^{\infty} \beta_j T^j + T^{r^{e-i+2}} \sum_{j=0}^{\infty} \beta_j^r T^{rj} = \frac{1}{x_{i-1}} + \left(\frac{1}{x_{i-1}}\right)^r = \frac{1}{x_i^r + x_i} = T^{r^{e-i+1}} \sum_{j=0}^{\infty} \alpha_j T^j.$$

By comparing the coefficients of $T^{j+r^{e-i+1}}$ in the above identity, we have that $\beta_0 = \alpha_0$ and $\beta_j$ can be easily computed recursively by the following formula for all $i \geq 1$.

$$\beta_j = \begin{cases} \alpha_j & \text{if } r \nmid j \\ \alpha_j - \beta_{j/r-1}^r & \text{if } r \mid j. \end{cases}$$

Therefore, the local expansion of $x_{i-1}$ at $P_\infty$ can be easily computed. □

As in the Hermitian case, we will actually need to index the messages of the code by the first $k$ coefficients $(f_0, f_1, \ldots, f_{k-1})$ of the local expansion of the function $f$ at $P_\infty$.

Let us define the local expansion map $\mathsf{ev}_{P_\infty} : \mathcal{L}((k + 2g_e - 1)P_\infty) \to \mathbb{F}_q^k$ that maps $f$ to $(f_0, f_1, \ldots, f_{k-1})$ where $f = T^{-(k+2g_e-1)}(f_0 + f_1 T + f_2 T^2 + \cdots)$ is the local expansion of $f$ at $P_\infty$.

**Claim 6.3.** $\mathsf{ev}_{P_\infty}$ *is an $\mathbb{F}_q$-linear surjective map. Further, we can compute $\mathsf{ev}_{P_\infty}$ using $\mathrm{poly}(k, g_e)$ operations over $\mathbb{F}_q$ given a representation of the input $f \in \mathcal{L}((k+2g_e-1)P_\infty)$ in terms of the basis (16).*

The proof of this claim is similar to Claim 3.3. Note that the kernel of $\mathsf{ev}_{P_\infty}$ is $\mathcal{L}((2g_e-1)P_\infty)$ which has dimension exactly $g_e$ by the Riemann-Roch theorem.

For each $(f_0, f_1, \ldots, f_{k-1}) \in \mathbb{F}_q^k$, we can therefore pick a pre-image in $\mathcal{L}((k + 2g_e - 1)P_\infty)$. For convenience, we will denote an injective map making such a unique choice by $\kappa_{P_\infty} : \mathbb{F}_q^k \to \mathcal{L}((k + 2g_e - 1)P_\infty)$. By picking the pre-images of a basis of $\mathbb{F}_q^k$ and extending it by linearity, we can assume $\kappa_{P_\infty}$ to be $\mathbb{F}_q$-linear, and thus specify it by a $(k + g_e) \times k$ matrix. We record this fact for easy reference below.

**Claim 6.4.** *The map $\kappa_{P_\infty} : \mathbb{F}_q^k \to \mathcal{L}((k+2g_e-1)P_\infty)$ is $\mathbb{F}_q$-linear and injective. We can compute a representation of this linear transformation using $\mathrm{poly}(k, g_e)$ operations over $\mathbb{F}_q$, and the map itself can be evaluated using $\mathrm{poly}(k, g_e)$ operations over $\mathbb{F}_q$.*

Now we redefine a version of the folded Garcia-Stichtenoth code that maps $\mathbb{F}_q^k$ to $(\mathbb{F}_q^m)^N$ by composing the folded encoding (17) from the original Definition 9 with $\kappa_{P_\infty}$.

**Definition 10** (Folded Garcia-Stichtenoth code using local expansion). *The folded Garcia-Stichtenoth code (FGS code for short)* $\mathsf{FGS}(N, k, q, e, m)$ *maps* $\mathbf{f} = (f_0, f_1, \ldots, f_{k-1}) \in \mathbb{F}_q^k$ *to* $\widetilde{\mathsf{FGS}}(N, k + 2g_e - 1, q, e, m)(\kappa_{P_\infty}(\mathbf{f})) \in (\mathbb{F}_q^m)^N$.

The rate of the above code equals $k/(Nm)$ and its distance is at least $N - (k + 2g_e - 1)/m$.

## 6.3 List decoding FGS codes

The list decoding part for the codes from the Garcia-Stichtenoth tower is almost identical to the Hermitian tower. We only sketch this part briefly.

If $f$ is a function in $\mathcal{L}((k+2g_e-1)P_\infty)$ whose encoding (17) agrees with the received word $\mathbf{y}$ in at least $t$ columns with $t > \frac{D+k+2g_e-1}{m-s+1}$ and

$$D = \left\lfloor \frac{N(m-s+1) - k + (s-1)g_e + 1}{s+1} \right\rfloor,$$

then there exist $A_i \in \mathcal{L}(DP_\infty)$ for $i = 1, 2, \ldots, s$ and $A_0 \in \mathcal{L}((D+k+2g_e-1)P_\infty)$ such that they are not all zero and

$$Q(f, f^{\sigma^{-1}}, \ldots, f^{\sigma^{-(s-1)}}) = A_0 + A_1 f + A_2 f^{\sigma^{-1}} + \cdots + A_s f^{\sigma^{-(s-1)}} = 0. \tag{18}$$

***Solving the functional equation for $f$.*** As in the Hermitian case, our goal next is to recover the list of solutions $f$ to the functional equation (18). Recall that our message functions lie in $\mathrm{Im}(\kappa_{P_\infty})$, so we can recover $f$ by recovering the top $k$ coefficients $(f_0, f_1, \ldots, f_{k-1})$ of its local expansion.

$$f = T^{-(k+2g_e-1)} \sum_{j=0}^{\infty} f_j T^j \tag{19}$$

at $P_\infty$. We now prove that $(f_0, f_1, \ldots, f_{k-1})$ for $f$ satisfying Equation (18) belong to a "periodic" subspace (in the sense of Definition 1) of not too large dimension.

**Lemma 6.5.** *The set of solutions $(f_0, f_1, \ldots, f_{k-1}) \in \mathbb{F}_q^k$ such that*

$$f = T^{-(k+2g_e-1)} \sum_{j=0}^{\infty} f_j T^j \in \mathcal{L}((k+2g_e-1)P_\infty)$$

*obeys equation*

$$A_0 + A_1 f + A_2 f^{\sigma^{-1}} + \cdots + A_s f^{\sigma^{-(s-1)}} = 0 \tag{20}$$

*when at least one $A_i$ is nonzero is an affine subspace $W$ of dimension at most $(s-1)\left\lceil \frac{k}{r-1} \right\rceil$.*

*Further, there are at most $q^{Nm+s+1}$ possible choices of the subspace $W$, each of which is $(s, r-1)$-periodic.*

*Given the representation of each $A_i$ w.r.t the basis (16), we can find a representation of $W$ in terms of the periodic subspace $U$ of dimension less than $s$, and the affine shifts in each window of $r-1$ coordinates, in the sense of Definition 1.*

*Proof.* Let $u = \min\{\nu_{P_\infty}(A_i) : i = 1, 2, \ldots, s\}$. Then it is clear that $u \leqslant 0$ and $\nu_{P_\infty}(A_0) \geqslant u - (k+2g_e-1)$. Each $A_i$ has a local expansion at $P_\infty$:

$$A_i = T^u \sum_{j=0}^{\infty} a_{i,j} T^j$$

for $i = 1, \ldots, s-1$ and $A_0$ has a local expansion at $P_\infty$:

$$A_0 = T^{u-(k+2g_e-1)} \sum_{j=0}^{\infty} a_{0,j} T^j$$

From the definition of $u$, one knows that the polynomial

$$B_0(X) := a_{1,0} + a_{2,0}X + \cdots + a_{s,0}X^{s-1}$$

is nonzero.

Assume that at $P_\infty$, the function $f$ has a local expansion (19). Then $f^{\sigma^{-i}}$ has a local expansion at $P_\infty$ as follows

$$f^{\sigma^{-i}} = \xi^{-(k+2g_e-1)i} T^{-(k+2g_e-1)} \sum_{j=0}^{\infty} \xi^{ji} f_j T^j,$$

where $\xi = 1/\gamma$.

The coefficient of $T^{d+u-(k+2g_e-1)}$ in the local expansion of $Q(f, f^{\sigma^{-1}}, \ldots, f^{\sigma^{-(s-1)}})$ is

$$0 = B(\xi^{d-(k+2g_e-1)}) f_d + \sum_{i=0}^{d-1} b_i f_i + a_{0,d}, \tag{21}$$

where $b_i \in \mathbb{F}_q$ is a linear combination of $a_{i,j}$ which does not involve $f_j$. Hence, $f_d$ is uniquely determined by $f_0, \ldots, f_{d-1}$ as long as $B(\xi^{d-(k+g_e-1)}) \neq 0$.

Let $S := \{0 \leq d \leq r-2 : B(\xi^{d-(k+g_e-1)}) = 0\}$. Then it is clear that $|S| \leq s-1$ since the order of $\xi$ is $r-1$ and $B_0(X)$ has degree at most $s-1$. Thus, $B(\xi^{d-(k+g_e-1)}) \neq 0$ if and only if $j \mod (r-1) \notin S$; and in this case $f_j$ is a fixed affine linear combination of $f_i$ for $0 \leq i < j$. Note that $B_0(X)$ has at most $(s-1)\left\lceil \frac{k}{r-1} \right\rceil$ roots among $\{\xi^i : i = 0, 1, \ldots, k-1\}$. It follows that the set of solutions $(f_0, f_1, \ldots, f_{k-1})$ is an affine space $W \subset \mathbb{F}_q^k$, and the dimension of $W$ is at most $(s-1)\left\lceil \frac{k}{r-1} \right\rceil$.

The fact that $W$ is $(s, r-1)$-smooth follows from (21) and noting that the coefficients $b_{d-j}$ for $j \geq 1$ in that equation are given by $B_j(\xi^{d-j-(k+2g_e-1)})$ where $B_j(X) := a_{1,j} + a_{2,j}X + \cdots + a_{s,j}X^{s-1}$. Therefore, once the values of $f_i$, $0 \leq i < (j-1)(r-1)$ are fixed, the possible choices for the next block of $(r-1)$ coordinates, $f_{(j-1)(r-1)}, \cdots, f_{j(r-1)-1}$, lie in an affine shift of a fixed subspace of dimension at most $(s-1)$. Further, the affine shift is an affine linear combination of the $f_i$'s in the previous blocks.

Finally, by the choice of $D$, the total number of possible $(A_0, A_1, \ldots, A_s)$ and hence the number of possible functional equations (20), is at most $q^{N(m-s+1)+s+1} \leq q^{Nm+s+1}$. Therefore, the number of possible candidate subspaces $W$ is also at most $q^{Nm+s+1}$. $\square$

Similar to the bound (13) for the Hermitian case, we conclude, after some simple calculations and using the upper bound on genus $g_e \leq r^e$, that one can find a representation of the $(s, r-1)$-periodic subspace containing all candidate messages $(f_0, f_1, \ldots, f_{k-1})$ in polynomial time, when the fraction of errors $\tau = 1 - t/N$ satisfies

$$\tau \leq \frac{s}{s+1}\left(1 - \frac{k}{N(m-s+1)}\right) - \frac{3m}{m-s+1}\frac{r^e}{mN}. \tag{22}$$

## 6.4 Combining FGS codes and h.s.e sets

Similarly to Section 5, we now show how to pre-code the messages of the FGS code with a h.s.e subset. The approach is similar, though we need one idea to ensure that we can pick parameters so that the base field $\mathbb{F}_q$ can be constant-sized and obtain a final list-size bound that is a constant independent of the code length. This idea is to work with a larger "period size" $\Delta$ for the periodic subspaces, based on the following observation.

**Observation 6.6.** *Let $W$ be an $(s, \Delta)$-periodic subspace of $\mathbb{F}_q^k$ for $k = b\Delta$. Then $W$ is also $(su, \Delta u)$-periodic for every integer $u$, $1 \leqslant u \leqslant b$.*

As in the Hermitian case, instead of encoding arbitrary $\mathbf{f} \in \mathbb{F}_q^k$ by the folded Garcia-Stichtenoth code (Definition 3), we will restrict the messages $\mathbf{f}$ to belong to the range of our h.s.e set. This will ensure that the affine space of solutions guaranteed by Lemma 6.5 can be efficiently pruned to a small list.

**Theorem 6.7.** *Let $r$ be a prime power, $q = r^2$, and $e \geqslant 2$ be an integer, and $\zeta \in (0,1)$. Let $k \leqslant q^{\zeta\Delta/2}$ be a positive integer. Let $\Delta \leqslant k$ be a multiple of $(r-1)$, say $\Delta = u(r-1)$ for a positive integer $u$.*

*Let $s, m$ be positive integers satisfying $1 \leqslant s \leqslant m \leqslant r-1$ and $s < \zeta r/12$. Finally let $N$ be an integer satisfying $k + 2r^e \leqslant Nm \leqslant (r-1)r^e$.*

*Consider the code $C_2$ with encoding $E_2 : \mathbb{F}_q^{(1-3\zeta)k} \to (\mathbb{F}_q^m)^N$ defined as*

$$E_2(\mathbf{x}) = \mathsf{FGS}(N,k,q,e,m)(\mathsf{HSE}(\mathbf{x})) ,$$

*for $\mathsf{HSE} : \mathbb{F}_q^{(1-3\zeta)k} \to \mathbb{F}_q^k$ from Definition 8 for a period size $\Delta$.*

*Then, with high probability over the choice of $\mathsf{HSE}$ with period size $\Delta$, this code has rate $R = (1-3\zeta)k/(Nm)$, can be encoded in $\mathrm{poly}(Nmq^{\zeta\Delta})$ time, and is $(\tau, \ell)$-list decodable in time $\mathrm{poly}(Nmq^{\zeta\Delta})$ for $\ell \leqslant O(1/(R\zeta))$ and*

$$\tau = \frac{s}{s+1}\left(1 - \frac{k}{N(m-s+1)}\right) - \frac{3m}{m-s+1}\frac{r^e}{mN} . \qquad (23)$$

*Proof.* This follows by just combining the ingredients we have developed so far. Since the genus $g_e$ is upper bounded $r^e$, the condition on $N, m$ meets the requirement for the construction of the folded codes based on Garcia-Stichtenoth tower in Definition 9.

Whp, the map $\mathsf{HSE}$ is well-defined and injective, and so $E_2$ is an injective encoding. The rate of the code is therefore clearly as claimed. By Theorem 4.6, Part 1, $\mathsf{HSE}$ can be computed in time $\mathrm{poly}(Nmq^{\zeta\Delta})$ and hence so can $E_1$ (as $\mathsf{FGS}$ is efficiently encodable as well).

The claimed value of the error fraction $\tau$ is just the bound (22). By Lemma 6.5, we know that the candidate messages found by the decoder lie in one of at most $q^{2Nm}$ possible $(s, r-1)$-periodic subspaces. By Observation 6.6, each of these subspaces is also $(su, \Delta)$-periodic. One can check that the conditions of Theorem 4.6 are met for our choice of $\zeta, q, \Delta, k$ and taking $su$ to play the role of $s$ (since $s < \zeta r/12$, we have $su < \zeta\Delta/10$).

Appealing to Theorem 4.6 and Lemma 4.7 with the choice $c = 2Nm/k = O(1/R)$, we conclude that there is a decoding algorithm running in time $\mathrm{poly}(Nmq^{\zeta\Delta})$ to list decode $C_2$ from a fraction $\tau$ of errors, outputting at most $O(1/(R\zeta))$ messages in the worst-case. □

Finally, all that is left to be done is to pick parameters to show how the above can lead to optimal rate list-decodable codes over a constant-sized alphabet which further achieve very good lists-size.

Let $\varepsilon > 0$ be a small positive constant, and a family of codes of length $N$ (assumed large enough) and rate $R \in (0, 1)$ is sought. Pick $n$ to be a growing parameter.

Let us pick $s = \Theta(1/\varepsilon)$, $m = \Theta(1/\varepsilon^2)$, $\zeta = \varepsilon/6$, $r = \Theta(1/\varepsilon)$, $q = r^2$, and $e = \lceil \frac{\log n}{\log r} \rceil$, $N = \lfloor \frac{(r-1)r^e}{m} \rfloor$, and $k = RNm(1+\varepsilon)$. This ensures that (i) there are at least $n = Nm$ rational places and so we get a code of length at least $n/m = N$, (ii) the rate of the code $C_2$ is at least $R$, and (iii) the error fraction (23) is at least $1 - R - \varepsilon$.

The remaining part is to pick a multiple $\Delta$ of $(r-1)$ so that the $k \leqslant q^{\zeta\Delta/2}$ condition is met. This can be achieved by choosing $u = \lceil \frac{\log n}{\log(1/\varepsilon)} \rceil$ and $\Delta = (r-1)u$. With these choices, we can conclude the following, which is the main final result of this paper.

**Theorem 6.8** (Main; Corollary to Theorem 6.7 with above choice of parameters). *For any $R \in (0, 1)$ and positive constant $\varepsilon \in (0, 1)$, there is a Monte Carlo construction of a family of codes of rate at least $R$ over an alphabet size $\exp(O(\log(1/\varepsilon)/\varepsilon^2))$ that are encodable and $(1-R-\varepsilon, O(1/(R\varepsilon)))$-list decodable in $\mathrm{poly}(N)$ time, where $N$ is the block length of the code.*

It may be instructive to recap why the Hermitian tower could not give a result like the above one. In the Hermitian case, the ratio $g_e/n$ of the genus to the number of rational places was about $e/r = e/\sqrt{q}$, and thus we needed $q > e^2$. Since the period $\Delta$ was about $q$, the running time of the decoder was bigger than $q^{\Omega(\zeta q)}$, whereas the length of the code was at most $q^{O(\sqrt{q})}$. This dictated the choice of $q \approx \log^2 n$, and then to keep the running time polynomial, we had to take $\zeta \approx (\log n \log \log n)^{-1}$.